\documentclass[12pt,preprint]{aastex}
\usepackage{graphicx}
\usepackage{amsmath, mathtools}
\usepackage{amssymb}
\usepackage{booktabs}  
\usepackage{color}

\def\eqp#1{(\ref{eq:#1})}
\def\eql#1{\label{eq:#1}}
\newcommand{\be}{\begin{equation}}
\newcommand{\ee}{\end{equation}}
\newcommand{\ba}{\begin{eqnarray}}
\newcommand{\ea}{\end{eqnarray}}
\def\order#1{\mathcal{O}\left(#1\right)}
\newcommand\bfu{\mathbf{u}}
\newcommand\bfx{\mathbf{x}}
\newcommand\bfr{\mathbf{r}}

\def\bfU{{\bf U}}
\def\cm{\mathrm{cm}}
\def\gm{\mathrm{gm}}
\def\secinv{\mathrm{s}^{-1}}

\def\rmM{\mathrm{M}}

\def\AU{\mathrm{AU}}
\newcommand\rma{\mathrm{a}}

\def\rmg{\mathrm{g}}
\def\rme{\mathrm{e}}
\def\rmf{\mathrm{f}}
\def\rmi{\mathrm{i}}
\def\rmp{\mathrm{p}}
\def\rms{\mathrm{s}}
\def\rmt{\mathrm{t}}
\def\rmK{\mathrm{K}}
\def\rmG{\mathrm{G}}
\newcommand\mm{\mathrm{mm}}
\def\phig{\phi_\rmg}
\def\phigo{\phi_{\rmg0}}

\def\Rcore{R_\mathrm{core}}

\def\Sigmag{\Sigma_\rmg}

\def\sigmatail{\sigma_\mathrm{tail}}
\def\sigmaG{\sigma_\rmG}

\def\epsg{\epsilon_\rmg}
\def\epsp{\epsilon_\rmp}
\def\epsgdd{\ddot{\epsilon}_\rmg}
\def\epspdd{\ddot{\epsilon}_\rmp}
\def\epsgd{\dot{\epsilon}_\rmg}
\def\epspd{\dot{\epsilon}_\rmp}

\def\rhog{{\rho_\rmg}}
\newcommand\Deltag{\Delta_\rmg}
\newcommand\Deltap{\Delta_\rmp}
\newcommand\rhogo{\rho_{\rmg 0}}
\newcommand\rhop{{\rho_\rmp}}
\newcommand\rhopo{\rho_{\rmp 0}}
\newcommand\rhot{\rho_\rmt}
\newcommand\rhoto{\rho_{\rmt 0}}

\newcommand\Stdyn{\mathrm{St}_\mathrm{dyn}}
\newcommand\Madyn{\mathrm{Ma}_\mathrm{dyn}}
\newcommand\Steta{\mathrm{St}_\eta}

\newcommand\tsedsing{t_\mathrm{sed,sing}}
\newcommand\ts{t_\rms}
\newcommand\tf{t_\rmf}
\def\tsinv{t_\rms^{-1}}

\def\tc{t_c}

\newcommand\tg{t_\rmg}

\def\ci{c_{\,\rmi}}

\newcommand\ceff{c_\mathrm{eff}}
\def\rb{\overline r}
\def\cbarEpstein{{\overline c}}

\newcommand\p{\partial}
\newcommand\tdyn{t_\mathrm{dyn}}

\newcommand\rbar{\overline{r}}
\newcommand\ppr{\frac{\p}{\p r}}
\newcommand\ppt{\frac{\p}{\p t}}

\newcommand\Jt{J_\mathrm{t}}
\newcommand\Jc{J_\mathrm{c}}

\newcommand\rg{r_\rmg}
\newcommand\rp{r_\rmp}
\newcommand\frg{f_\mathrm{rg}}

\newcommand\rfront{r_\mathrm{front}}
\newcommand\tarrival{t_\mathrm{arrival}}
\newcommand\tcollapse{t_\mathrm{collapse}}

\newcommand\Msol{M_\odot}

\newcommand\Ri{\mathrm{Ri}}
\newcommand\rcore{r_\mathrm{core}}

\newcommand\sprime{{\scriptscriptstyle\prime}}
\newcommand\rprime{{r^\sprime}}
\newcommand\varpimax{\varpi_\mathrm{max}}
\newcommand\StK{\mathrm{St}_\mathrm{K}}

\newcommand\Reff{R_{0,\mathrm{eff}}}
\newcommand\epsfilter{\epsilon_\mathrm{filter}}

\shorttitle{Spherical collapse of solids}
\shortauthors{Shariff \& Cuzzi}

\begin{document}
\title{The Spherically Symmetric Gravitational Collapse of a Clump of Solids in a Gas}
\author{Karim Shariff and Jeffrey N. Cuzzi}
\affil{NASA Ames Research Center, Moffett Field, CA 94035}
\slugcomment{Astrophys. J., Accepted, March. 3, 2015}

\begin{abstract}
In the subject of planetesimal formation, several mechanisms have been identified that create dense particle clumps in the solar nebula.  The present work is concerned with the gravitational collapse of such clumps, idealized as being spherically symmetric.  Fully nonlinear simulations using the two-fluid model are carried out (almost) up to the time when a central density singularity forms.  We refer to this as the collapse time.  The end result of the study is a parametrization of the collapse time, in order that it may be compared with timescales for various disruptive effects to which clumps may be subject in a particular situation.  An important effect which determines the collapse time is that as the clump compresses, it also compresses the gas due to drag.  This increases gas pressure which retards particle collapse and can lead to oscillation in the size and density of the clump.  In the limit of particles perfectly coupled to the gas, the characteristic ratio of gravitational force to gas pressure becomes relevant and defines a two-phase Jeans parameter, $\Jt$, which is the classical Jeans parameter with the speed of sound replaced by an effective wave speed in the coupled two-fluid medium.  The parameter $\Jt$ remains useful even away from the perfect coupling limit because it makes the simulation results insensitive to the initial density ratio of particles to gas ($\Phi_0$) as a separate parameter.  A simple ordinary differential equation model is developed.  It takes the form of two coupled non-linear oscillators and reproduces key features of the simulations.  Finally, a parametric study of the time to collapse is performed and a formula (fit to the simulations) is developed.  In the incompressible limit $\Jt \to 0$, collapse time equals the self-sedimentation time which is inversely proportional to the Stokes number.  As $\Jt$ increases, the collapse time decreases with $\Jt$ and eventually becomes approximately equal to the dynamical time.  Values of collapse time versus clump size are given for a minimum-mass solar nebula.  Finally, the timescale of clump erosion due to turbulent strain is estimated.

\end{abstract} 

\section{Context}

Once planetesimals, solid bodies between 10 and 100 km in size, Kelvin--Helmwere formed in the solar system, the growth to larger bodies is believed to have been straightforward, and the mechanisms that have been put forward succeed in forming planetary embryos \citep{Wetherill_and_Stewart1989, Kokubo_and_Ida1998, Goldreich_etal2004a}.  Likewise, the growth from micron- to cm-sized particles can take place by sticking.  At this point, particle collisions induced by their relative velocities lead to mostly to fragmentation \citep{Brauer_etal2008} or bouncing and impact compaction \citep{Zsom_etal2010} and further growth is stalled.  Calculations of coagulation, which either solve the Smoluchowski equation or employ Monte Carlo methods, have yet to incorporate special features of turbulence that could be important in the coagulation process.  For instance, particles that are concentrated by turbulence have the smallest relative velocities \cite[Fig. 4 in][]{Zaichik_and_Alipchenkov2003}, which reduces the collision rate but also reduces bouncing and shattering.  Current coagulation calculations indicate that a barrier exists at cm size, and researchers have sought to overcome it by positing mechanisms that concentrate particles; these mechanisms are thought to be followed by self-gravity driven contraction.  That body of work is reviewed in more detail in the next section.  

The present work considers one of the ways in which self-gravity driven contraction is frustrated by the presence of gas, in particular by its pressure.  Although dust is itself pressure-less (Brownian motion is unimportant for the particle masses of interest here), it indirectly feels the gas pressure.  This is because as the particles compress due to self-gravity, they also partially compress the gas via the drag force.  This increase in gas pressure causes the gas to resist compression and this in turn causes particles to resist compression \citep{Cuzzi_etal2008}.  The purpose of the present work is to quantify this effect and how it is relieved as the Stokes number increases, i.e., as drag becomes weaker.  To this end, the simple situation of an initially spherical clump of particles embedded in a gas initially at rest is considered.  In addition to the one-dimensional (radial) simulations, a simple ODE model is developed which captures the essential features of the simulations.  The main end product is a formula (fit to the simulations) for collapse time as a function of the three governing parameters: a two-phase Jeans number $\Jt$ (which is introduced in \S\ref{sec:general}), the Stokes number, and the initial density ratio $\Phi_0$ of solids to gas. 

\section{Review of particle concentration mechanisms}

\subsection{Midplane dynamics}

A layer of particles can settle toward the disk midplane if this region of the disk is sufficiently free of turbulent velocity fluctuations, as may happen \citep{Gammie1996} when there is insufficient ionization to sustain magneto-rotational instability.  A rich set of mechanisms operate in this layer, which we now discuss.

Seeking an alternative to coagulation, \cite{Safronov1969} and \cite{Goldreich_and_Ward1973} showed that in the absence of gas, the midplane layer undergoes (axisymmetric) gravitational instability at a sufficiently short radial wavelength.  \cite{Sekiya1983} included the presence of gas assuming it is ``perfectly coupled to particles'', meaning that the particle velocity $\bfU$ exactly equals the gas velocity $\bfu$.  This becomes true (modulo a gravitationally induced sedimentation velocity) as particles become very small.  Sekiya showed that in this limit, gas pressure imposes a very high critical value (e.g., $6 \times 10^6$ at 5 AU) for the ratio ($\Phi \equiv \rhop/\rhog$) of particle to gas density required for gravitational instability.  This is easily understood, as described near Equation \eqp{Phi_crit} below, by replacing the sound speed with the effective sound-speed in the coupled two-fluid medium.  This replacement was suggested by Safronov (\citeyear{Safronov1987}, p.~137)
\footnote{The two-phase sound speed for liquid-gas mixtures is used in the context of lunar formation by \cite{Thompson_and_Stevenson1983}.} and rediscovered by Cuzzi et al. (\citeyear{Cuzzi_etal2008}, p.~1435).  Sekiya finds that another mode exists whose critical particle density assumes a Roche-like value of $\rho_\rmp^* = 0.6 \Msol / \varpi^3$, where $\varpi$ is the orbital radius.  This density leads to a much lower critical value of $\Phi = 170$ at $\varpi = 5$ AU.  It is important to note that this mode is incompressible, i.e., it cannot increase the particle density.  While such a mode can lead to fragmentation of the midplane layer, it can contribute to planetesimal formation only via (slow) sedimentation towards the center of the fragments.  How the constraint imposed by gas pressure relaxes with increasing Stokes number has yet to be quantified for this situation, to our knowledge.

Even if one accepts that the incompressible mode is relevant to planetesimal formation, the midplane layer cannot easily achieve the required density due to self-generated turbulence.  Particles at the midplane revolve faster than the gas which moves slower than the Keplerian speed because it has slight pressure support. This results in an Ekman-like layer with vertical shear which makes it susceptible to turbulence driven by the Kelvin--Helmholtz instability, which could prevent the critical density from being achieved \citep{Weidenschilling1980}.   On the other hand, the layer can be stably stratified due to the presence of particles.  The ratio of stabilizing buoyancy to shear is called the Richardson number, $\Ri$.  Ignoring differential rotation together with the Coriolis force, and assuming that particles are perfectly coupled to the gas, one has classical stratified shear flow which is known to be stable provided that the local $\Ri \geq 1/4$ everywhere \citep[][p.~328]{Drazin_and_Reid1981}.  \cite{Garaud_and_Lin2004} find that as the particle layer settles, this (sufficient) criterion for flow stability is violated before Sekiya's incompressible gravitational mode becomes unstable.  \cite{Cuzzi_etal1993} solved the Reynolds-averaged two-fluid equations for a turbulent midplane layer using mixing-length-type models for various covariances in the more general case of finite stopping time.  They conclude that Sekiya's critical particle density is approached at 10 AU only for particles of about 1 m in size.  Likewise, the critical density is not achieved in the simulations of \cite{Johansen_etal2006}, which use Lagrangian particles and ignore Keplerian shear.  The latter can reduce the strength of the Kelvin--Helmholtz vortices by tilting them away from the vertical shear; however, it can also transiently strengthen them by stretching.  Simulations by \cite{Barranco2009} included the Coriolis force and radial shear in the single-fluid limit (i.e., assuming that the particle velocity exactly equals the gas velocity) and concluded that the flow is turbulent.  However, these simulations could not obtain the equilibrium particle density distribution because, in the perfectly coupled limit, the particle mass fraction is an advected scalar and retains its maximum value from the initial condition.  This is where work on midplane Kelvin--Helmholtz turbulence currently stands.

An interesting mechanism known as the streaming instability was discovered by \cite{Youdin_and_Goodman2005}.  It creates particle concentrations in the midplane layer and under certain circumstances can counteract particle diffusion due to the Kelvin--Helmholtz instability.  The basic state is the \cite{Nakagawa_etal1986} solution for radial and azimuthal particle drift relative to the gas flow. \cite{Youdin_and_Goodman2005} perform a local axisymmetric linear stability analysis for this basic state assuming uniform gas and particle density.  In particular, vertical shear leading to Kelvin--Helmholtz instability is absent.  Non-linear simulations with the same set-up are presented in \cite{Johansen_and_Youdin2007}.  The streaming instability does not depend on self-gravity and simultaneously clumps particles both radially ($x$) and vertically ($z$), i.e., unstable modes have $k_x \neq 0$ and $k_z \neq 0$; see Figure 2 in \cite{Youdin_and_Goodman2005}.  Clumping in the radial direction is explained by \cite{Jacquet_etal2011} as follows: Imagine a perturbation mode in which particles are radially compressed.  If the stopping time is sufficiently short, the particles will also compress the gas via drag, leading to a local pressure maximum.  The pressure maximum then acts as a further attractor for particles (as is well known and described below), which closes the feedback loop.  Note that the analysis of \cite{Youdin_and_Goodman2005} takes the gas to be incompressible; in this limit the increase in gas pressure arises from enforcement of the incompressibility constraint.  For the mechanism of radial clumping put forward by \cite{Jacquet_etal2011}, one expects optimal instability growth at an intermediate particle size small enough for particles to drag the gas but large enough to undergo radial drift.  Finally, we mention two computational studies \citep{Johansen_etal2009, Bai_and_Stone2010} whose set-up allows for both the Kelvin--Helmholtz and streaming instabilities to occur.  These studies also include multiple particle sizes in each simulation and therefore have the additional effect of differential radial drift.  They conclude that significant clumping takes place when the (vertically averaged) density ratio of dust to gas is supersolar and particles sufficiently large.

Since the Kelvin--Helmholtz and streaming instabilities work against each other in concentrating particles, one may ask under what conditions is the latter sufficiently strong that gravitational instability can take place.  In a brief computational study, \cite{Johansen_etal2009} introduced particles with four radii, namely 3, 6, 9, and 12 cm at 5 AU.  They found that when nebular dust to gas ratio was $Z = 0.01$, clumping was insignificant.  However at a super-solar value of $Z = 0.02$, clumping sharply increased. 

\subsection{Trapping in pressure highs}

A local maximum of gas pressure at orbital radius $\varpi = \varpimax$ attracts solids towards it \citep{Haghighipour_and_Boss2003}.  This is because for $\varpi < \varpimax$,  we have $\p p/\p \varpi > 0$  leading to faster than Keplerian gas velocity.  This boosts the orbital radius of particles, while the opposite is true for particles at $\varpi < \varpimax$.  A more quantitative rendition of this is as follows.  The analysis of  \cite{Nakagawa_etal1986} is easily extended to non-small stopping times to give for the radial drift velocity of particles:
\be
   U_\varpi = -2\eta u_\rmK \left[\frac{\StK}{(1 + \Phi) + \StK^2}\right], \eql{drift_speed}
\ee
where $u_\rmK$ is the Keplerian velocity, $\Phi = \rho_\rmp/\rho_\rmg$ is the particle loading, $\StK = t_\rms u_\rmK/\varpi$ is the Stokes number based on the characteristic orbital time, and $\eta$ is the pressure gradient parameter
\be
   \eta = -\frac{\varpi}{2\rho_\rmg u_\rmK^2}\frac{\p p}{\p \varpi}.
\ee
From \eqp{drift_speed}, we see that the radial drift velocity is maximized for $\StK = (1 + \Phi)^{1/2} > 1$.  At 3 AU in a minimum-mass nebula this implies particles with radii $a \gtrsim 70$ cm.  In other words, this mechanism is most effective for meter-sized bodies.  Related to the above phenomenon is the trapping of particles by large-scale anti-cyclonic vortices \citep{Barge_and_Sommeria1995}, which are pressure highs, or in the vortices created by magneto-rotational turbulence \citep{Johansen_etal2007}.  Meter-sized bodies are also the ones that drift towards the central star most rapidly (which is, of course, not coincidental) and Johansen et al. suggest that such bodies can avoid this fate by being trapped in vortices where they can gravitationally collapse on dynamical timescales.  This suggestion neglects fragmentation of the boulders in the eddies by collision.

\subsection{Turbulent concentration}

The difficulty with the above mechanisms is that they require either large particles or a nebula with an enhanced solids-to-gas mass fraction.  Another scenario for planetesimal formation \citep{Cuzzi_etal2008, Cuzzi_etal2010} envisions a direct path from chondrule (mm) sized particles to asteroidal sized bodies.  It invokes the phenomenon of turbulent concentration, first articulated by Maxey (\citeyear{Maxey1987}, p.~458), whereby particles centrifuge out from regions of high vorticity and accumulate in regions of high strain.  Simulations of homogeneous isotropic turbulence confirmed this \citep{Squires_and_Eaton1991} and found that the effect was most pronounced for particles having a certain ratio of particle stopping time.  Subsequently, \cite{Wang_and_Maxey1993} refined this result by finding that particles having Stokes number $\Steta \approx \ts / t_\eta \approx 1$ are the ones most concentrated.  Here $\ts$ is the particle stopping or response time and $t_\eta$ is the Kolmogorov time.  It so happens that chondrule-sized particles approximately fulfill this condition \citep{Cuzzi_etal2001}.  The larger and slower eddies concentrate larger particles \citep{Bec_etal2007, Pan_etal2011}.  The current status of this effort will be provided in the concluding remarks.


\subsection{The present work in context}

Many of the above mechanisms produce dense particle clumps.  They include the Kelvin--Helmholtz instability \citep{Johansen_etal2006}, preferential concentration in turbulence \citep[e.g.][]{Cuzzi_etal2001, Cuzzi_etal2008, Cuzzi_etal2010, Pan_etal2011}, and the streaming instability \citep{Johansen_and_Youdin2007, Bai_and_Stone2010}.
The present work is concerned with the gravitational collapse of such clumps.  Our collapse solutions cover the relevant regimes of the three non-dimensional parameters, namely, the Stokes number, particle loading, and the two-phase Jeans number.  In reality, a particle clump will also be subject to dispersive effects from within and without.  For the case of preferential concentration by turbulence, the former includes turbulent gas motions within the clump which lead to a particle dispersion velocity.  The latter includes straining by eddies of the same size and larger than the clump, and ram gas pressure because dense clumps revolve faster than the gas \citep{Cuzzi_etal2008}. 
While these dispersal mechanisms should be studied in more detail in the future, in \S\ref{sec:erosion} we provide an estimate for the rate of erosion by turbulent strain.

\section{Formulation}\label{sec:analysis}

\subsection{Governing Equations}

The equation of motion of a single particle of mass $m_\rmp$ and radius $a$ subject to gas drag is
\be
   m_\rmp \frac{d\bfU}{dt} = \frac{4\pi a^2}{3} \,\cbarEpstein \rhog \left(\bfu - \bfU\right), \eql{eom}
\ee
where $\bfU$ is the particle velocity, $\bfu$ is the gas velocity, and $\rhog$ is the gas density.  The quantity $\cbarEpstein$ is the mean thermal velocity of gas molecules defined by \cite{Epstein24} as
\be
   \cbarEpstein = \left(8/\pi\right)^{1/2} \ci, \eql{cbarEpstein}
\ee
where $\ci$ is the isothermal sound speed.
Equation \eqp{eom} uses the \cite{Epstein24} drag law, which is valid when $\left| \bfu - \bfU\right| \ll \ci$ and $a < \lambda_\rmg$, where $\lambda_\rmg$ is the mean free path of the gas.  We will employ the two-fluid treatment for solid particles, which applies mass and momentum conservation to a differential volume containing many particles.  Let all particles be identical (the so-called mono-disperse case); this is convenient but not essential.  For a differential volume, consider a physical volume large enough to contain many particles but small enough that $\bfu(\bfx)$ does not change appreciably across it.  In a protoplanetary disk the distance between particles ($\sim 1$ m) is much smaller
than the smallest scale across which changes in velocity occur (the Kolmogorov scale $\sim 1$ km).  We also require that $\bfU$ not change appreciably across the volume and be a single-valued function of particle position.  This assumption disallows ``crossing trajectories'' such as when two particle streams traveling in different directions and/or at different velocities interpenetrate without colliding.  In this case, the continuum treatment averages the particle velocity in the interpenetration zone leading to unphysical behavior.  One way to treat this case in the future is to solve a kinetic equation for the particle number density in the position-velocity phase space \citep{Chalons_etal2012}.
The final assumption is that particles are sufficiently separated from each other that the drag law \eqp{eom}, which is for a single particle in an infinite medium, remains valid.

When solid particles are treated as a continuum phase, the following two-fluid equations of motion result:
\ba
\rhog_{,t} + (\rhog u_i)_{,i} & = & 0,
\hskip 0.5truecm \mathrm{(gas\ mass)} \eql{fmass}\\
(\rhog u_i)_{,t} + (\rhog u_i u_j)_{,j} & = & - p_{,i} + \rhog g_i - d_i,
\hskip 0.5truecm \mathrm{(gas\ momentum)} \eql{fmom}\\
e_{,t} + \left[u_j (p + e)\right]_{,j} &=& \rhog g_i u_i - d_i u_i - q_{j,j}
\hskip 0.5truecm \mathrm{(gas\ energy)} \eql{fenergy}\\
\rhop_{,t} + (\rhop U_i)_{,i} & = & 0,
\hskip 0.5truecm \mathrm{(particle\ mass)} \eql{pmass}\\
(\rhop U_i)_{,t} + (\rhop U_i U_j)_{,j} & = & \rhop g_i + d_i.
\hskip 0.5truecm \mathrm{(particle\ momentum)} \eql{pmom}
\ea
Here $\rhop$ is the density of the particle phase, $q_j$ is the heat flux vector, and
\be
   d_i = \beta \cbarEpstein \rhog\rhop\left(u_i - U_i\right), \mathrm{\ with\ }\beta \equiv \frac{4\pi a^2}{3 m_\rmp},
\ee
is the drag force per unit volume exerted on the particle phase by the fluid phase.
The quantity $g_i = \varphi_{,i}$ is the gravitational acceleration, where the potential $\varphi$ satisfies Poisson's equation
\be
    \varphi_{,ii} = - 4 \pi G (\rhog + \rhop).
\ee
The equation of state for an ideal gas of specific heat ratio $\gamma$ is:
\be
   e = \frac{p}{\gamma - 1} + \frac{1}{2} \rhog |\bfu|^2. \eql{state}
\ee
We have implemented the adiabatic and isothermal cases but have performed calculations for only the latter, in which case the equations of energy \eqp{fenergy} and state \eqp{state} are replaced by
\be
    p = \rhog {\ci}^2.
\ee
For the adiabatic case, the heat flux vector is set to zero ($q_i = 0$).

\subsection{General considerations concerning the role of gas pressure}\label{sec:general}

While the competition between gas pressure and gravity is readily appreciated in the star-formation context, it is less obvious in the context of gravitational collapse of solid particles.  In this sub-section we introduce a two-phase acoustic speed $\ceff$ which was first appreciated by \cite{Sekiya1983}; see also the review article by Marble \citeyear{Marble1970}, p.~403) on dusty gases.  This leads to a two-phase Jeans parameter, $\Jt$, which is the classical Jeans parameter with the acoustic speed replaced by $\ceff$.

Let the particle response time $\ts$ be much shorter than the flow timescale $\tf$ over which the velocity field changes following a gas particle; $\tf$ can be obtained from the velocity gradient tensor.  Then, particles and gas approximately follow each other and we may adopt the \cite{Maxey1987} expansion
\be U_i = u_i + \ts F_i(\bfx, t) + \order{\ts^2}, \eql{Maxey} \ee
where $F_i(\bfx, t)$ remains to be determined.  In writing \eqp{Maxey} we have assumed that each quantity has been non-dimensionalized, in particular, that time has been non-dimensionalized using $\tf$.  Substituting \eqp{Maxey} into the particle momentum equation in acceleration form,
\be \frac{\p U_i}{\p t} + U_j \frac{\p U_i}{\p x_j} = g_i + (u_i - U_i)/\ts, \ee
gives
\be U_i = u_i + \ts \left(g_i - \frac{Du_i}{Dt}\right) + \order{\ts^2}.  \eql{generalized_sedimentation} \ee
The second term on the right of \eqp{generalized_sedimentation} is simply a sedimentation velocity generalized to include a D'Alembert term due to gas acceleration.  Here we need only the leading order result, $U_i = u_i$, which allows one to add the momentum equations for the gas and particles to give (for the isothermal case)
\begin{align}
{\rhot}_{,t} + (\rhot U_i)_{,i} &=  0, \eql{cmass}\\
(\rhot U_i)_{,t} + (\rhot U_i U_j)_{,j} &=  - \ci^2 (\phig \rhot)_{,i} + \rhot g_i \eql{cmom}
\end{align}
where $\rho_\rmt \equiv \rhog + \rhop$ is the total density and $\phig \equiv \rhog/\rhot$ is the gas mass fraction.  Note that the particle loading $\Phi\equiv\rhop/\rhog=\phig^{-1}-1$ and conversely $\phig = (1 + \Phi)^{-1}$.  One may then define a two-phase Jeans $\Jt$ parameter as the ratio of gravitational to pressure force terms in \eqp{cmom}:
\be
   \Jt = \frac{\rho_\rmt G \ell^2}{\phig \ci^2},
\ee
where $\phi_\rmg$ and $\ell$ are the characteristic gas mass fraction and size of the region.  The criterion for collapse on the dynamical timescale $(\rhot G)^{-1/2}$ is
\be \Jt > \Jc, \eql{Jc} \ee
where $\Jc$ is a critical value of $\order{1}$.  The ratio of characteristic free-fall velocity $R_0(\rhot G)^{1/2}$ to $\ci$ is the dynamical Mach number $\Madyn$ so 
\be
   \Jt = (1 + \Phi) \Madyn^2.
\ee
The criterion \eqp{Jc} then becomes
\be
   \Madyn > \left[(1 + \Phi)^{-1}\Jc\right]^{1/2} \approx (1 + \Phi)^{-1/2}. \eql{Madyn}
\ee
In other words, the gas must have (sufficient) compressibility to allow dynamical collapse of particles that are tightly coupled to it.  Otherwise, the best that one may expect is particle sedimentation through an incompressible gas.

To be complete, Equations \eqp{cmass} and \eqp{cmom} need to be supplemented with a transport equation for $\phig$.  We have
\begin{align}
   \frac{1}{\rhop}\frac{D\rhop}{Dt} &= \nabla\cdot\bfU,\eql{pmass2}\\
   \frac{1}{\rhog} \frac{D\rhog}{Dt} &= \nabla\cdot\bfU,\eql{fmass2}
\end{align}  
where
\be
\frac{D}{Dt} \equiv \frac{\p}{\p t} + \bfU\cdot\nabla.
\ee
Subtracting \eqp{fmass2} from \eqp{pmass2} gives
\be
   \frac{D}{Dt}\left(\ln\Phi\right) = 0.
\ee
Thus $\ln\Phi$ is a passive scalar and so is any function of it, in particular $\phig$:
\be
   \frac{D\phig}{Dt} = 0 \eql{phig}
\ee
This means that in the perfectly coupled limit, the maximum value of the particle loading cannot change from its initial value.
If, in addition, $\phig$ is initially uniform\footnote{A study of acoustics in a dusty medium where $\phig$ is not initially uniform and where the Stokes number ranges from very small to order unity values would be an interesting exercise.} it remains so and may be removed from under the pressure gradient in \eqp{cmom}.  The equations of motion then become completely equivalent to those of a compressible isothermal gas with effective speed of sound
\be   
   \ceff = \ci \sqrt{\phig}. \eql{ceff}
\ee
The role of the effective speed $\ceff$ seems to have been first recognized by \cite{Safronov1969}, \cite{Marble1970}, and Sekiya (\citeyear{Sekiya1983}, p.~1122) who accounts for the presence of gas in the Goldreich-Ward midplane instability.  The equivalence \eqp{ceff} allows many known results for isothermal gases to be directly translated to a perfectly coupled particle-gas system.  Here are three examples: (i) Jeans' criterion for gravitational collapse translates to
\be
   \lambda > \left(\frac{\pi \phig\ci^2}{G\rhot}\right)^{1/2},
\ee
for instability where $\lambda$ is the wavelength of a disturbance.
(ii) The dispersion relation for an infinitesimally thin Keplerian gas sheet subject to axisymmetric disturbances
is (in the local shearing sheet approximation):
\be
   \omega^2 = k^2 \ci^2 + \Omega_\rmK^2 - 2\pi G\Sigmag k.
\ee
To adapt this to a perfectly coupled particle-gas sheet, one makes the replacements $\ci^2 \to \ci^2 \phig$ and $\Sigma_\rmg \to \Sigma_\rmp$.  A range of unstable wave numbers exists when
\be
   \phig < \left(\frac{\pi G \Sigma_\rmp}{\Omega_\rmK \ci}\right)^2. \eql{phig_crit}
\ee
For a minimum mass nebula \citep{Hayashi1981} we have $\ci = 10^5 \varpi^{-1/4} \ \cm\ \secinv$, $\Sigma_\rmp = 7.1 \varpi^{-3/2}$ $\ \rmg \ \cm^{-2}$, and $\Omega_\rmK = 2 \times 10^{-7} \varpi^{-3/2}\ \secinv$, where $\varpi$ is the orbital radius in AU.  For $\varpi = 1$ AU
one obtains
\be
   \Phi > 1.8 \times 10^8, \eql{Phi_crit}
\ee
which is clearly a very tall order and should be compared with the criterion $\Phi > 2.5 \times 10^8$ that Sekiya (his Table IIa) obtains using a more complex analysis.  
(iii) \cite{Shu77} presented a solution for the rate of infall during self-similar collapse of a singular isothermal gas sphere.  If instead one had a gas-particle sphere with an initial total density distribution $\rhot \sim r^{-1}$ and uniform gas mass fraction, then with the replacement $\ci \to \ci \sqrt{\phig}$, Shu's result becomes
\be
   \dot{M} = .975 \ci^3 \phig^{3/2} / G.
\ee

With the above discussion motivating the use of $\ceff$ and $\Jt$, we now describe numerical calculations of the collapse of a spherical clump of particles without a restriction to small stopping times.

\subsection{Governing equations for spherical collapse}

With spherical symmetry the governing equations have the form:
\begin{align}
\frac{\p\rhog}{\p t} + \frac{1}{r^2}\ppr\left(r^2 \rhog u\right) &= 0,\eql{a}\\
\ppt\left(\rhog u\right) + \frac{1}{r^2}\ppr\left(r^2 \rhog u^2\right) &= -\frac{\p p}{\p r} + \rhog g - \beta\cbarEpstein\rhop\rhog(u-U), \eql{b}\\
\frac{\p\rhop}{\p t} + \frac{1}{r^2}\ppr\left(r^2 \rhop U\right) &= 0,\eql{c}\\
\ppt\left(\rhop U\right) + \frac{1}{r^2}\ppr\left(r^2 \rhop U^2\right) &= \rhop g + \beta\cbarEpstein\rhop\rhog(u-U),\eql{d} \\
\frac{\p e}{\p t} + \frac{1}{r^2}\ppr\left[r^2\left(e + p\right)u\right] &= \rhog g u - u\beta\cbarEpstein\rhop\rhog(u-U) 
- \frac{1}{r^2}\ppr\left(r^2 q_r\right).\eql{e}
\end{align}
Here $u$ and $U$ denote the radial velocity of the gas and particles, respectively, and $q_r$ is the radial radiative heat flux.  The gravitational acceleration is
\be
g(r) = -\frac{G M(r)}{r^2} \mathrm{\ where\ } M(r) =\int_0^r 4\pi \rprime^2 (\rhog(\rprime) + \rhop(\rprime) - \rhogo)\, d\rprime. \eql{gravity}
\ee
Note that the density $\rhogo$ of the background gas has been subtracted out in $M(r)$ to ensure that the background gas by itself remains static.  This is equivalent to assuming that there is a background pressure gradient that balances the gravity of the background gas.  In the present work, we shall assume that compressive and frictional heating are so slow compared with radiative transfer that the gas remains isothermal.

\subsection{Numerical scheme and validation}

Spatial derivatives on the left hand sides of Equations \eqp{a}--\eqp{e} are computed using a second-order accurate finite-volume scheme.  Grid points (where the time-evolved quantities reside) are located at the midpoints of computational shells in the radial direction.  In a finite-volume scheme, the rate of change of a shell-averaged quantity is the difference of fluxes at the shell faces divided by the shell volume.  At second-order, the distinction between a shell-centered and shell-averaged quantity is immaterial; recall that the midpoint rule for quadrature is second-order accurate.  We first compute fluxes per unit area (i.e., without the $r^2$ factor) at grid-points, linearly interpolate them to shell faces, multiply by $r^2$, take the difference, and divide by the shell volume.

The origin requires a slightly special treatment and two schemes are implemented.
In the first, a grid-point is placed at the origin
and fluxes per unit area are linearly interpolated on to the spherical surface surrounding it.  These fluxes determine the rates of change of densities and energy at the origin.  The rates of change of radial gas and particle velocities at the origin are explicitly set to zero (as required by spherical symmetry).  All results presented this work are obtained using the first scheme.
In the second scheme, the origin is taken to be a flux location rather then a grid point.  Fluxes can be interpolated there using the symmetry or anti-symmetry of various quantities across the origin.  The only non-zero ``flux'' at the origin is the pressure since it is both an even function across the origin and not multiplied by an $r^2$ factor.  

At the outer radial boundary, a choice between two boundary conditions for the gas is provided.  The first is a non-reflective boundary condition \citep{Thompson1987} for the gas-dynamic Euler terms which sets the rate of change of incoming characteristic variables to zero.  All spatial derivatives required to implement this condition are calculated using one-sided differences.  The second choice is as follows.
When the inflow velocity of the gas at the outer boundary is subsonic, which was always the case for the collapse runs, there is one incoming characteristic (for isothermal flow where the characteristic speeds are $u+c$ and $u-c$) and one may specify one quantity.  We chose to set the gas density equal to its ambient value.  All the collapse runs were run with this boundary condition.  Some tests were conducted to assess differences when the non-reflective boundary condition was used instead.  The differences were very small, indicating that the location of the outer boundary was sufficiently far.   For the particle equations, a characteristic decomposition is not possible and the fluxes required on the outer surface of the last shell are obtained by linear extrapolation.  

\begin{figure}
\begin{center}
\includegraphics[width=5.5in]{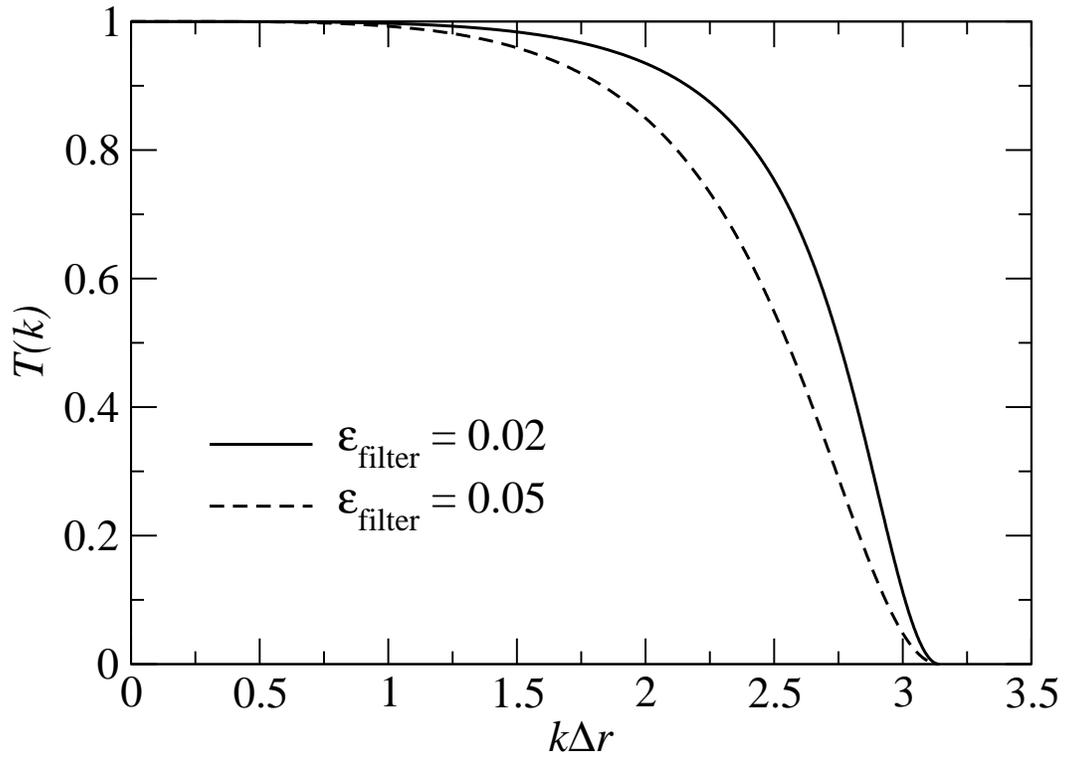}
\end{center}
\caption{Transfer function $T(k)$ for the numerical filter in the range of $\epsfilter$ values used in the collapse runs.  $k$ is the wavenumber and $\Delta r$ is the grid size.}
\label{fig:tf}
\end{figure}

Unlike shock-capturing schemes,  the scheme just described is free of numerical dissipation.  As a result, there is no mechanism to damp oscillations at the Nyquist wavelength which equals twice the grid spacing.  Such waves are numerically generated, even in a linear problem, whenever there are inhomogeneities such as changes in mesh spacing, boundary treatments, or changes in coordinate Jacobians.  These waves have zero phase velocity, and once created, remain.  Centrally differenced schemes therefore require a small amount of filtering.  At the end of every time step we applied the fourth-order Pad\'e filter given by Equation (C.2.1) in \cite{Lele1992} with his coefficients $\beta = d = 0$.  It has a transfer function that is near unity and drops off sharply near the Nyquist wavenumber $k\Delta r = \pi$ where $\Delta r$ is the grid spacing.  The strength and sharpness of the filter is determined by a parameter $\epsfilter$ such that Lele's $\alpha = 1/2 - 2\epsfilter$.  A value of $\epsfilter = 0$ gives no filtering.  The collapse runs used values of $\epsfilter$ between 0.02 and 0.05; Figure~\ref{fig:tf} plots the transfer function of the filter for these two values.

The gravitational acceleration at each grid-point is obtained by assuming the density factor $\rhog(r) + \rhop(r) - \rhogo$ to be piecewise linear in $r$ and analytically performing the integration in \eqp{gravity} for $M(r)$.  Time advancement is performed using a fourth-order Runge-Kutta scheme.

\begin{figure}
\begin{center}
\includegraphics[width=4.5in]{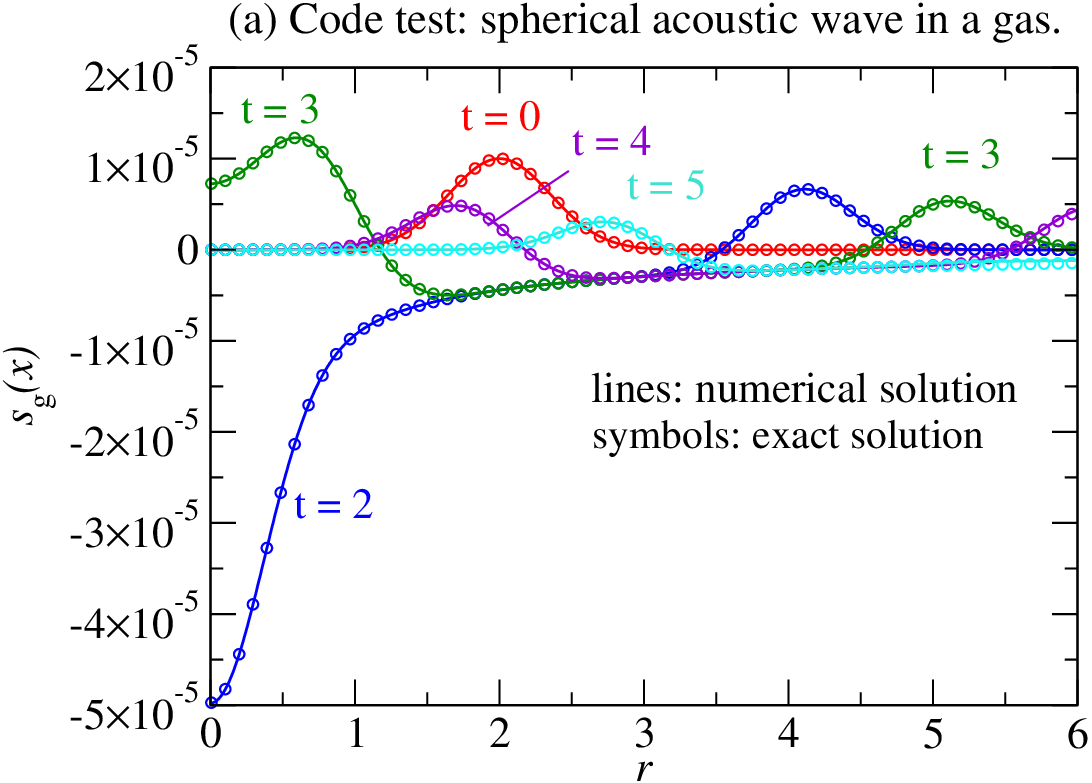}
\vskip 0.75truecm
\includegraphics[width=4.5in]{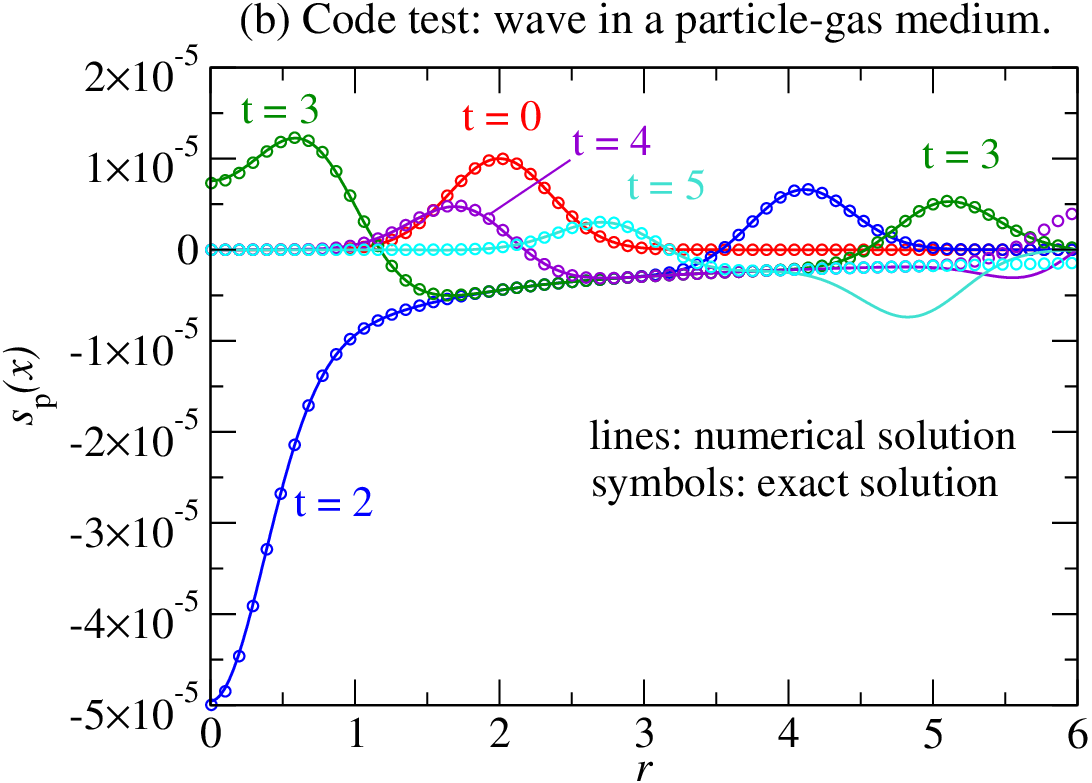}
\end{center}
\caption{Code validation tests.  (a) Acoustic wave in a gas.  The ordinate $s_\rmg(r) \equiv\rhog(r)/\rhogo - 1$ is the relative perturbation in gas density.  Sound speed $\ci = 1$.
(b) Acoustic wave in a particle-gas medium.  Stopping time $\ts = 0.001$.  $s_\rmp(r) \equiv \rhop(r)/\rhopo-1$ is the relative perturbation in particle density.  Particle loading $\Phi = 3$ which gives a gas mass fraction $\phig = 1/4$.  We chose $\ci = 2$ to give $\ceff = \ci \phig^{1/2} = 1$.  1000 grid points.}
\label{fig:code_test}
\end{figure}

The purely gas dynamic terms in the code, together with the non-reflective condition at the outer boundary, were validated against the exact solution (starting from arbitrary initial conditions) for spherical acoustic waves beautifully described by \citet[\S279]{Rayleigh1945} accounting for reflection at the origin.  The initial condition we used was a Gaussian bump for the density and radial velocity perturbation,  made symmetric and anti-symmetric, respectively, about the origin.
Figure~\ref{fig:code_test}a shows the result of using the first treatment of the origin (where the origin is a grid-point).  The simulation (lines) is in excellent agreement with the exact solution (symbols).  To avoid Nyquist oscillations in the acoustic test, we found that the parameter, $\epsilon_\mathrm{filter}$, governing the strength of the filter and its cut-off wavenumber needed to be set to $0.03$.  The second scheme (where the origin is a flux surface) required a lower $\epsilon_\mathrm{filter} = 0.01$ in the acoustic test.  In the few comparisons we made for collapse cases, both schemes required comparable smoothing.  The largest value used for $\epsilon_\mathrm{filter}$ in the collapse runs was $0.05$.

Quadrature for evaluating the gravitational acceleration \eqp{gravity} was validated using an analytical ($\cos^2(kr)$) density profile.
To test the implementation of the particle transport and mutual drag terms, we performed a calculation for a particle-gas acoustic wave in the tightly coupled limit with uniform gas mass fraction ($\phig$) initially.  We would like to suggest this as a test case for other two-fluid codes as well.  In this limit (see \S\ref{sec:general}), we should have the same exact solution as a gaseous acoustic wave except that the speed of sound is replaced by $\ci\sqrt{\phig}$.  The stopping time was set to $\ts = 0.001$ and the gas mass fraction was set to $\phig = 1/4$.  To obtain a solution with the same functional form as for the gaseous acoustic wave described in the last paragraph, the speed of sound $\ci$ was doubled.  Figure~\ref{fig:code_test}b shows good agreement between the simulation (lines) and exact solution (symbols).  However, comparing the cyan line and symbols, we see that there is some spurious reflection of the wave at the outer boundary ($r = 6$).  In the problem of gravitational collapse of particles considered later in this work, the outer region of the domain is free of particles and so this issue does not arise.

For the collapse runs,
the mesh consists of a uniformly and finely spaced inner region where the density gradient is strongest.  This is followed by a second region where the mesh is stretched.  Finally, there is a third region where the mesh is uniform again.

\section{Results}

\subsection{Initial condition}

The initial clump consists of a core of radius $\rcore$ having uniform particle density, and a Gaussian tail of length $\sigmatail$.  The particle clump is embedded in a gas of initially uniform density $\rhogo$:
\be
   \rhop(r, 0) = \rhopo \begin{dcases}
                                1, & r \leq \rcore\\
                                 \rme^{-(r - \rcore)^2/\sigmatail^2}& r > \rcore.\\
                                  \end{dcases}, \hskip0.5truecm \rhog(r, 0) = \rhogo. \eql{ic}
\ee
The choice of a uniform particle density in the core makes the gas-particle wave speed uniform (in the tightly coupled limit) and therefore the wave dynamics is easier to analyze.  Furthermore, this choice is in accord with the assumptions of the ODE model.  The necessity for a having a smooth tail is purely numerical: because we do not use shock-capturing methods, discontinuities are not allowed.  
The size of the tail is set to $\sigmatail = 0.2 \rcore$ and the initial clump radius is defined as the sum $R_0\equiv \rcore + \sigmatail$.  
A comparison with a purely Gaussian $\rhop(r, 0)$ profile will be made in \S\ref{sec:Gaussian}.
The code requires computation of the particle velocity $U$ from the momentum $\rhop U$.  To avoid division by very small numbers a minimum value of $10^{-6}$ was used for $\rhop(r, 0)$.  The outermost radius of the computational domain is $r_\mathrm{max} = 4$ and the total number of mesh points is $\approx 1500$.
The evolution starts from rest.
Corresponding to initial conditions at the origin we define the dynamical (i.e., free-fall) time and the particle stopping time:
\be
   \tdyn \equiv \left(\rhopo G\right)^{-1/2}, \hskip 0.5truecm
   \ts \equiv \left(\beta \cbarEpstein \rhogo \right)^{-1}. \eql{timedefs}
\ee
The initial condition is characterized by three non-dimensional parameters defined using conditions at the origin: (i) The particle loading
\be
   \Phi_0 \equiv \rhop(0, 0)/\rhogo. \eql{Phidef}
\ee
We will often instead use the gas mass fraction $\phigo = (\Phi_0 + 1)^{-1}$. 
(ii) The two-phase Jeans parameter
\be
   \Jt \equiv \frac{\rhoto G R_0^2 (1 + \Phi_0)}{{\ci}^2}, \eql{Jtuniform}
\ee
where $\rhoto = \rhopo + \rhogo$.  Since the wave period 
\be
   t_\mathrm{wave} = R_0 / \ci \phigo^{1/2},
\ee
we have
\be
   \Jt = (t_\mathrm{wave} / \tdyn)^2.
\ee
(iii) The Stokes number based on dynamical time:
\be
   \Stdyn \equiv \ts/\tdyn.
\ee
Corresponding to the dimensions of mass, length, and time we may set any three quantities to unity in the code.  For these we chose $R_0$, $\ci$, and $\rhogo$.  The quantity we use to assess the progress of collapse is the radial centroid of particle density:
\be
   \rbar(t) \equiv \frac{\int_0^\infty r \rhop(r)\, dr}{\int_0^\infty \rhop(r)\, dr}. \eql{rbdef}
\ee

\subsection{First simulation} 
\label{sec:vary_Jeans}

\begin{figure}
\begin{center}
\includegraphics[width=5.5in]{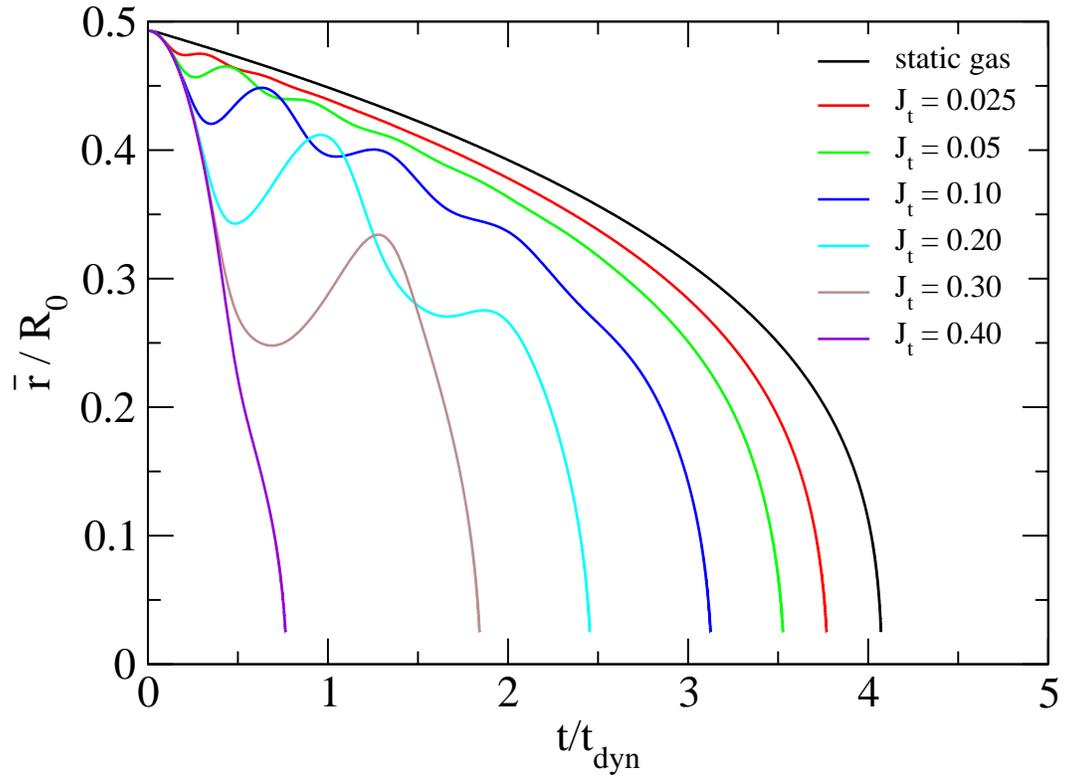}
\end{center}
\caption{Radial centroid $\rbar(t)$ of the particle density distribution $\rhop(r)$ for various two-phase Jeans numbers $\Jt$ ($\Phi_0 = 100$, $\Stdyn = 0.02$).}
\label{fig:rb_vary_Jt_Phi100_ts0.02}
\end{figure}

The first simulation was for the case $\Phi_0 = 100$ and $\Stdyn = 0.02$.
Figure~\ref{fig:rb_vary_Jt_Phi100_ts0.02} plots the time evolution of the radial particle centroid $\rb(t)$ for various Jeans parameters $\Jt$.  As $\Jt \to 0$ the gas becomes incompressible and with spherical symmetry this implies that it must become static.  Indeed, as $\Jt \to 0$ the curves tend to the behavior (black line) for a static gas, which was computed by setting $u = 0$ in the code.  This is the incompressible self-sedimentation limit and will be discussed in more detail later.  Monotonic collapse on the dynamical timescale occurs for $\Jt = 0.40$ (violet line).  For intermediate values of $\Jt$ a combination of bouncing and sedimentation is observed prior to collapse.  The first bounce occurs due to the propagation of an expansion wave (in both the gas and particle phases) from the particle-lean exterior to the particle-rich interior.  The time at which the first minimum in $\rb(t)$ occurs is consistent with wave propagation at the two-phase acoustic speed $\ceff = \phig^{1/2}\ci$.  The wave is then reflected from the origin as a compression, propagates outward to the edge of the particle core where it reflects as a rarefaction, and so on.  This sets up an oscillating standing wave whose behavior (in the absence of gravity) will be discussed in more detail in \S\ref{sec:standing}.

\subsection{The self-sedimentation limit} \label{sec:sedlimit}

We saw in the previous sub-section that, in the incompressible limit $\Jt \to 0$, the gas remains static with uniform density $\rhogo$.  In addition, consider the case where the local ratio of stopping time to dynamical time is everywhere small, i.e.,
\be
\frac{t_\mathrm{s}}{t_\mathrm{dyn}(r)} = \frac{(G \rhop(r))^{1/2}}{\beta \cbarEpstein \rhogo} \ll 1. \eql{local_Stokes}
\ee
Note that this condition will always fail in the inner core in the very late stages of collapse. 
If condition \eqp{local_Stokes} holds then particles will be close to terminal velocity 
where gravity and drag terms balance each other, acceleration being negligible in comparison.
The terminal velocity is then given by
\be
   U_\mathrm{term}(r) = g(r) \ts.\eql{Uterm}
\ee
The mass $M_S$ enclosed by any collapsing material surface $S$ is constant, which makes the
gravity at the surface, $g_S \propto M_S/r_S^2$, increase as the surface collapses.  Hence the sedimentation
velocity accelerates with time.
It was verified that for the static gas simulation in \S\ref{sec:vary_Jeans}, the particle velocity is closely given by 
\eqp{Uterm} except at late times in the inner core.  The particle dilatation associated with \eqp{Uterm} is
\be
   \Delta_\rmp \equiv \nabla \cdot \bfU =  - \alpha \rhop,
   \mathrm{\ where\ } \alpha \equiv 4\pi G \ts.
\ee 
The mass conservation equation $D\rhop/Dt = -\rhop \Delta_\rmp$ can then be solved in the (assumed) uniform density particle core to give the evolution of particle density:
\be
   \rhop(t) = \rhopo \left(1 - \alpha \rhopo t\right)^{-1}. \eql{rhop_sed}
\ee
According to \eqp{rhop_sed}, the volume of the clump ($\propto \rhop^{-1}$), not its radius, decreases linearly with time.
A singularity, which may be termed the sedimentation singularity, occurs at 
\be
   \tsedsing = (\alpha \rhopo)^{-1} = (4 \pi \Stdyn)^{-1} \tdyn, \eql{t_sed_sing}
\ee
which will be referred to as the sedimentation time.
Note that previous estimates \citep{Cuzzi_etal2008, Cuzzi_etal2010} do not account for the acceleration in sedimentation velocity.  As a result they overestimate the sedimentation time by a factor of $\pi$.

\begin{figure}
\begin{center}
\includegraphics[width=5.5in]{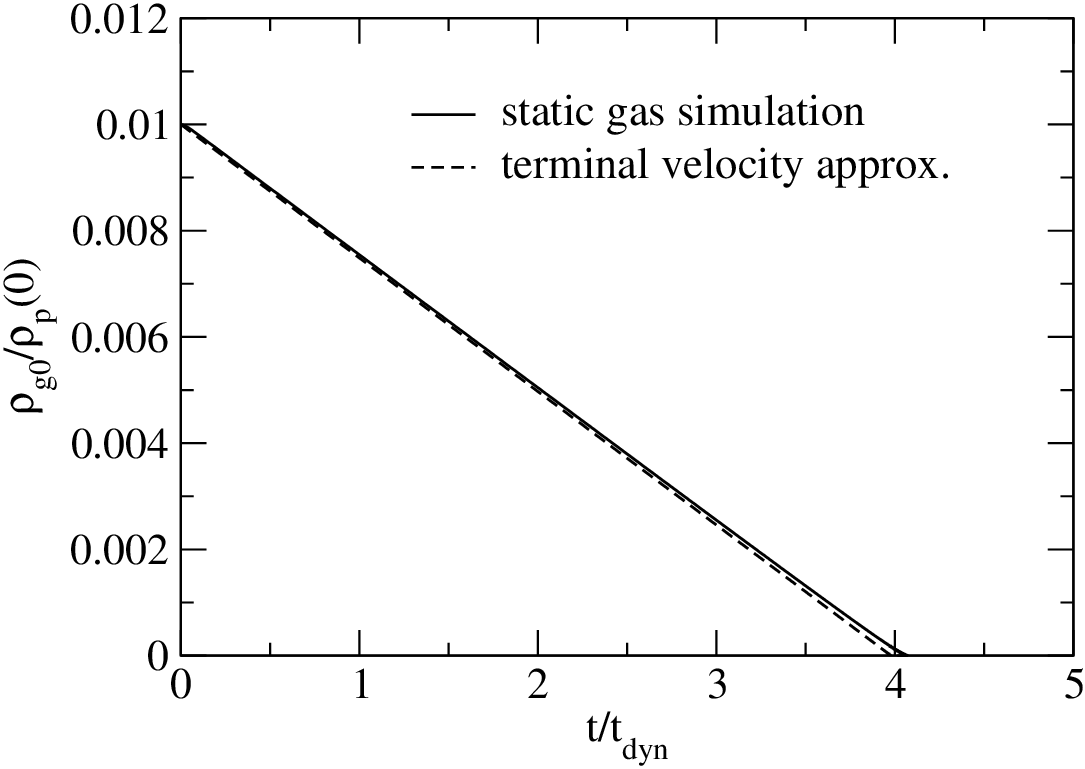}
\end{center}
\caption{Normalized specific volume at the origin comparing the result of the static gas simulation with equation \eqp{rhop_sed}. ($\Phi_0 = 100$, Stokes number $\Stdyn = 0.02$).}
\label{fig:specific_volume_sed}
\end{figure}

Figure~\ref{fig:specific_volume_sed} shows that the behavior predicted by \eqp{rhop_sed} agrees well with the static gas simulation at $\Stdyn = 0.02$. 
Obviously, the result \eqp{t_sed_sing} cannot hold indefinitely as $\Stdyn$ increases since the collapse time can never be smaller than $\tdyn$.  Deviations for larger $\Stdyn$ will be studied in \S\ref{sec:collapse_time}.

\subsection{Ordinary differential equation model}
 
Here we develop a simple model for the behavior of the clump near the origin (for the isothermal case).  To leading order, the dependent variables behave as follows near the origin:
\begin{align}
   \rhog(r,t) &= \rhog(t) + \order{r^2},\eql{forma}\\
   \rhop(r,t) &= \rhop(t) + \order{r^2},\eql{formb}\\
   u(r,t) &= \Deltag(t) r /3 + \order{r^3},\eql{formc}\\
   U(r,t) &= \Deltap(t) r/3 + \order{r^3}\eql{formd},
\end{align}
where $\Deltag(t)$ and $\Deltap(t)$ denote gas and particle dilatations, respectively, at the origin.  Note that a simplified notation is being used in which an argument of $t$ alone (or no argument at all) indicates the value at the origin.  One substitutes the forms \eqp{forma}--\eqp{formd}  into the governing equations \eqp{a}--\eqp{d}, equates coefficients of $r^0$ in the mass conservation equations, and coefficients of $r^1$ in the momentum equations.  However, note that to get the $r^1$ coefficient of the pressure gradient requires the density to quadratic order; a similar closure problem would present itself no matter how high the order of the expansion.  Therefore, the pressure gradient term, which will require a model, is left in its exact form for now.  The result is the ODE system:
\ba
\frac{d\rhog}{dt} &=& - \rhog \Deltag, \eql{ode1}\\
\frac{d\Deltag}{dt} &=& -\frac{1}{3} \Deltag^2 - 4\pi G (\rhop + \rhog - \rhogo) + \beta\cbarEpstein \rhop\left(\Deltap - \Deltag \right)
        + \Pi(t), \eql{ode2}\\
\frac{d\rhop}{dt} &=& - \rhop \Deltap, \eql{ode3}\\
\frac{d\Deltap}{dt} &=& -\frac{1}{3} \Deltap^2 - 4\pi G (\rhop + \rhog - \rhogo) - \beta\cbarEpstein \rhog\left(\Deltap - \Deltag \right). \eql{ode4}
\ea
In equation \eqp{ode2} the pressure gradient term (exact) is:
\be
   \Pi(t) \equiv - \left[\frac{3}{\rhog(r, t) r}\frac{\p p}{\p r}\right]_{r = 0}, \eql{pg_term}
\ee
where $p(r, t) = \rhog(r, t) \ci^2$ for the isothermal case.  Two models for $\Pi(t)$ will be introduced.  The first one (subscript M1) is
\be
   \Pi(t) \approx \Pi_{\rmM1}(t) = - \frac{3(\rhogo - \rhog(t)) \ci^2}{\rhog(t) \frg^2 \rg^2(t)}, \eql{pg_model}
\ee
where $\frg$ is a modeling constant.  The quantity $\rg(t)$ is the radius of the region of compressed gas and is obtained by mass conservation:
\be
   \rg(t) = \left[\rhogo / \rhog(t)\right]^{1/3} \rg(0), \eql{rg_model}
\ee
with its initial value set equal to the initial radius of the particle clump, i.e., $\rg(0) = R_0$.  The model constant $\frg$ in Equation \eqp{pg_model} was calibrated to $\frg = 0.43$ by obtaining the closest match to one simulation case ($\Phi_0 = 100, \Jt = 0.3, \Stdyn = 0.02$).  
Figure~\ref{fig:specific_volume}a shows the the time evolution of the inverse particle density at the origin as given by the simulations for various $\Jt$, while Figure~\ref{fig:specific_volume}b shows the corresponding predictions of the model.  
The ODE model reproduces many features of the simulations, however, the depth of collapse preceding the first bounce is smaller.  It was suspected that this difference arises because the pressure gradient model \eqp{pg_model} fails to account for the time delay in arrival of the pressure wave at the origin. 
This is confirmed in Figure~\ref{fig:pg_term} which shows that the actual (solid line) pressure gradient term $\Pi(t)$ at the origin begins to rise after a delay compared to the model (dashed) expressions \eqp{pg_model} and \eqp{rg_model}.  The second model allows for such a delay by tracking a wavefront location $\rfront(t)$ as follows:
\be
   \frac{d \rfront(t)}{dt} = u(\rfront, t) - \ceff, \eql{rfront}
\ee
with an initial location set to the particle core boundary (defined in equation \ref{eq:ic}):
\be
   \rfront(0) = \rcore.
\ee
The total wave speed on the right of \eqp{rfront} is that of the inward propagating characteristic with the gas speed given from \eqp{formc}
\be
   u(\rfront, t) =  \Deltag(t) \rfront /3,
\ee
and wave speed (relative to the gas) given by
\be
   \ceff(t) = \left(\frac{\rhog(t)}{\rhog(t) + \rhop(t)}\right)^{1/2} \ci.
\ee
Let $\tarrival$ be the time that the wave arrives at the origin.  Then the pressure gradient term is turned on only for $t > \tarrival$, i.e.,
\be
   \Pi(t) \approx \Pi_\mathrm{delay}(t) = H(t - \tarrival) \Pi_{\rmM1}\left(t\right), \eql{Pidelay}
\ee
where $H(t)$ is the Heaviside function.  $\Pi_\mathrm{delay}(t)$ constitutes our second model.
Calibration for the same case as before ($\Jt = 0.3, \Phi_0 = 100, \Stdyn = 0.02$) gave $\frg = 0.4579$ for the model constant.  The extra digits of precision result from the fact that we tried our best to nail the $\Jt = 0.3$ case, which lies in a sensitive region of the parameter space with a very deep compression followed by a bounce.  Figure~\ref{fig:specific_volume}c shows that the ODE model with arrival delay gives a better reproduction of the time and depth of the initial compression, however, the bounces are higher than the simulations for $\Jt \gtrsim 0.3$.

It may be noted that the ODE model can be written in a form bearing an analogy to that of two coupled (non-linear) oscillators.  For a velocity field of the form \eqp{formc} the acceleration of a gas particle is:
\be
   \frac{d^2\rg}{dt^2} = \frac{Du}{Dt} = \frac{\p u}{\p t} + u\frac{\p u}{\p r} = \frac{\rg}{3} \left(\frac{d\Deltag}{dt} + \frac{\Deltag^2}{3}\right), 
\ee
and similarly for a solid particle.  Hence \eqp{ode2} and \eqp{ode4} become
\begin{align}
   \frac{d^2\rg}{dt^2} &= -\frac{4\pi}{3} G \left(\rhop + \rhog - \rhogo\right) \rg + \beta\cbarEpstein\rhop\left(U - u\right) + \frac{\Pi(t)\rg}{3},\eql{osca}\\
   \frac{d^2\rp}{dt^2} &= -\frac{4\pi}{3} G \left(\rhop + \rhog - \rhogo\right) \rp - \beta\cbarEpstein\rhog\left(U - u\right). \eql{oscb}
\end{align}
The densities on the right of these equations are the following functions of $\rg$ and $\rp$ due to mass conservation:
\be
   \rhop(t) = \left(\frac{\rp(0)}{\rp(t)}\right)^3\rhop(0), \hskip 0.5truecm \rhog(t) = \left(\frac{\rg(0)}{\rg(t)}\right)^3\rhog(0).
\ee
The terms in \eqp{osca} and \eqp{oscb} that are proportional to $\rg$ and $\rp$ represent spring-like (non-linear) restoring forces and the drag terms provide damping.

\subsection{Standing wave without gravity}
\label{sec:standing}

To develop insight into the nature of bouncing oscillations, gravity was activated for only a short initial period $t/\tdyn \le 0.01$ to provide an initial velocity,  after which it was switched off.  Figure~\ref{fig:standing_wave}a shows profiles of gas density for one period of the standing wave.

Figure~\ref{fig:standing_wave}b shows that the ODE model (dashed line) reproduces the period and damping rate of density fluctuations at the origin quite well.  This gives us confidence to use it to develop further insight.  Linearizing the ODE model by substituting
\be
   \rg(t) = \rg(0) \left[1 + \epsg(t) + \order{\epsg^2}\right], \hskip 0.5truecm \rp(t) = \rp(0) \left[1 + \epsp(t) + \order{\epsp^2}\right]
\ee
into the oscillator equations \eqp{osca} and \eqp{oscb} (without the gravity terms and considering times larger than the pressure gradient arrival time) gives to leading order:
\begin{align}
   \epsgdd &= \tsinv \Phi_0 \left(\epspd - \epsgd\right) - \omega^2 \epsg, \eql{epsg}\\
   \epspdd &= - \tsinv \left(\epspd - \epsgd\right) \eql{epsp}.
\end{align}
The frequency
\be
   \omega \equiv \left(\frac{3\ci^2}{\frg^2 \rp^2(0)}\right)^{1/2},
\ee
is inversely proportional to the sound crossing time and arises from the pressure gradient model.  The system of equations \eqp{epsg} and \eqp{epsp} have eigenvalues $\lambda_m$ ($m = 0,\ldots,3$) each of which corresponds to a mode with time behavior $\exp(\lambda_m t)$.  The first eigenvalue is zero and the other three are roots of the cubic polynomial
\be
   \lambda^3 + \tsinv \phigo \lambda^2 + \omega^2 \lambda + \omega^2 \tsinv = 0.
\ee
In working out the roots of the cubic, the non-dimensional parameter
\be
   \varepsilon \equiv \left(\phigo \omega \ts\right)^2
\ee
emerges.  It involves the stopping time and will later be seen to set the damping rate of oscillations.  It can be expressed as
\be
   \varepsilon = \frac{3}{\frg^2} \phigo^2 \left(\frac{\ts}{\tc}\right)^2,
\ee
where $\tc = \rp(0) / \ci$ is the sound crossing time.  For the example depicted in Figure~\ref{fig:standing_wave} we have $\varepsilon \sim 10^{-3}$.
To leading order in $\varepsilon$ and the initial gas mass fraction $\phigo$, the non-zero eigenvalues are
\begin{align}
   \lambda_1       &= - \omega \varepsilon^{-1/2}, \eql{lambda1}\\
   \lambda_{2,3} &= \omega
                                   \left[-\frac{1}{2} \varepsilon^{1/2} \pm i \left(\phigo - \varepsilon/4\right)^{1/2}\right]. \eql{lambda23}
\end{align}
The first eigenvalue gives a rapidly damped non-oscillating mode.  If 
\be 
   \phigo > \varepsilon/4, \eql{cond} 
\ee
then the second two eigenvalues correspond to a damped oscillation as realized in Figure~\ref{fig:standing_wave}.  This is analogous to the underdamped case for a harmonic oscillator.  The oscillation frequency is given by
\be
   \omega_\mathrm{osc} = \omega \left(\phigo - \varepsilon/4\right)^{1/2}.
\ee
Without the $\varepsilon/4$ term the result corresponds to the frequency of waves propagating at speed $c_\mathrm{eff}$.  The $\varepsilon/4$ term gives a frequency correction due to damping analogous to that for a damped spring-mass system.
If condition \eqp{cond} is not satisfied then all three modes are damped without oscillation.  This is analogous to the overdamped case of the harmonic oscillator.  We expect that condition \eqp{cond} will also roughly determine whether bouncing oscillations will be present in the presence of gravity.  For example for $\Jt = 0.20$ and $\Phi_0 = 100$, Equation \eqp{cond} predicts that bouncing should disappear for $\Stdyn > .13$.  Indeed, we found no bouncing in the simulations for $\Stdyn = 0.15$.

For the case of Figure~\ref{fig:standing_wave}, Equation \eqp{lambda23} gives an oscillation period of $0.53\tdyn$ and an $\rme^{-1}$ damping period of $0.70\tdyn$.  The corresponding values in the simulation are $0.59\tdyn$ and $0.96\tdyn$, respectively.

\subsection{Collapse time across the parameter space}
\label{sec:collapse_time}

Collapse time $\tcollapse$ was recorded in the code as the instant when the normalized density centroid $\rbar/R_0$ falls below 0.03.  The plot of $\rb(t)$ becomes very nearly vertical at this time and so the actual instant of the singularity $\rbar/R_0 \to 0$ is very close to the recorded value.

We begin by investigating dependence on $\Stdyn$, fixing the particle loading at $\Phi_0 = 100$.  Figure~\ref{fig:t_collapse100}a shows that on average (i.e., apart from oscillations in slope, which become quite pronounced for $0.30 \lesssim\Jt \lesssim 0.40$) the curves fan out linearly with respect to $\Stdyn^{-1}$ from the point $(\Stdyn^{-1} = 1.0, \tcollapse/\tdyn = 0.57)$.  The dashed lines in Figure~\ref{fig:t_collapse100}a represent a simple linear (with cut-off) formula \eqp{formula} to be developed below.  Figures~\ref{fig:t_collapse100}b and \ref{fig:t_collapse100}c show that both versions of the ODE model provide reasonably good quantitative predictions.  Figure~\ref{fig:t_collapse10} shows the simulation result for a smaller particle loading of $\Phi_0 = 10$.  Comparing it with Figure~\ref{fig:t_collapse100}a one concludes that while details of the slope oscillations change, the overall behavior is insensitive to $\Phi_0$ in the range $10 \le \Phi_0 \le 100$ (keeping $\Jt$ fixed).  One also notes that in the incompressible (static gas) limit $\Jt \to 0$ (solid black line), the terminal velocity approximation $\tcollapse/\tdyn = (4\pi\Stdyn)^{-1}$ (red dashed line) is good to within 10\% for $\Stdyn < 0.05$.  

Next, the dependence on $\Jt$ is investigated; see Figure~\ref{fig:t_collapse_over_t_dyn}.  Again, there are oscillations in slope as well as a few cases where increasing the strength of gravity (i.e., $\Jt$) actually increases the collapse time.  This is a consequence of bouncing: for these cases, increasing $\Jt$ causes a deeper initial collapse but the bounce is even stronger, which increases the time to collapse.  Apart from such anomalies and oscillations, the behavior is linear.  We write an estimate for the collapse time by drawing a straight line from the terminal velocity approximation at the left end ($\Jt = 0, \tcollapse/\tdyn = \max[(4\pi \Stdyn)^{-1}, 0.65]$) to the point $(\Jt = 0.43, \tcollapse/\tdyn = 0.65)$.  The max function has been introduced in writing the coordinates of the left end-point in order to prevent the collapse time from dropping below $0.65 \tdyn$ when the terminal velocity approximation breaks down for large $\Stdyn$.   
The straight line is given by
\be
F(\Jt) = 
\max[\left(4 \pi \Stdyn\right)^{-1}, 0.65]\left[1 - \Jt/0.43\right] + (0.65/0.43) \Jt.
\ee
Finally we write
\be
   \frac{\tcollapse}{\tdyn} = \max(F(\Jt), 0.65), \eql{formula}
\ee
to keep the collapse time from going below (0.65 times) the dynamical time for large $\Jt$.
Figure~\ref{fig:t_collapse_over_t_dyn} shows formula \eqp{formula} using dashed lines.
The reader may also refer back to Figures \ref{fig:t_collapse100}a and \ref{fig:t_collapse10} and the dashed lines therein to assess the performance of formula \eqp{formula} as a function of $\Stdyn^{-1}$.
Comparing the results for $\Phi_0 = 10$ and 100 one concludes again that the dependence on particle loading $\Phi_0$ as a separate parameter is weak and Equation \eqp{formula} continues to work quite well.  In other words, the presence of $\Phi_0$ in the definition $\Jt$ captures most of the influence of $\Phi_0$.

Figure~\ref{fig:t_collapse_versus_Jt_ODE} shows the predictions of collapse time given by both versions of the ODE model.  The results of the simulations, previously shown in Figure~\ref{fig:t_collapse_over_t_dyn}, are depicted in gray.  The overall agreement is satisfactory for both models.  The model without delay gives a better prediction for smaller $\Jt$ while the model with delay is slightly more accurate for 
$\Jt \gtrsim 0.4$.

\subsection{Development of particle density profiles versus $r$}

Time evolution of the particle density profile was studied for the following set of twelve cases: $\Phi_0 \in \left\{10, 100\right\} \times \Stdyn \in \left\{0.005, 0.05\right\} \times \Jt \in \left\{0, 0.2, 0.4\right\}$; see Figure \ref{fig:profiles}.  Profiles are compared at the same value of the ``progress variable'' $\rb(t)$ (the density centroid defined in Equation \ref{eq:rbdef}).  Insets are used to show $\rb(t)$.  Since $\rb(t)$ sometimes oscillates in time, it is necessary to state that profiles are recorded the first time that $\rb(t)$ goes below the given value.
Figures \ref{fig:profiles}a--c illustrate the $\Jt$ dependence, other parameters being kept fixed.  For $\Jt = 0$ (incompressible static gas), the profile remains flat in the collapsing core.  This is in accord with the solution obtained in \S \ref{sec:sedlimit} where the terminal velocity approximation was invoked.  Slight departure from this behavior at the final instant is likely due to breakdown of this approximation.  Note that as $\Jt$ increases (Figures \ref{fig:profiles}a--c), the profiles become less flat in the core.  In particular, when $\Jt$ is increased from $0.2$ to $0.4$, the knee in some profiles (e.g., for $\rb/R_0 = 0.25$, violet) disappears.  In all the other cases this was the only difference observed between the $\Jt = 0.2$ and $0.4$ cases.  Hence, we omit all other plots for $\Jt = 0.4$.

Next compare Figures \ref{fig:profiles}b and d; in the latter case $\Phi_0$ is ten times larger.  Except for a factor of ten difference in the ordinate, the profiles in the two plots are nearly the same.  This was observed to be true for all the other cases in the set.

Between Figures \ref{fig:profiles}d and e, $\Stdyn$ has been increased by ten.  Despite the large difference in the collapse time, profiles at the same value of the progress variable are nearly identical.  This conclusion holds for all the other cases in the set.  

To determine the functional form of the particle density profile about the origin when the singularity forms there, we plot in Figure \ref{fig:profiles}e profiles for all twelve cases on log-log axes at the final instant of the simulations, when $\rb/R_0 = 0.03$.  A power law region develops with a slope slightly steeper than $-2$.

\subsection{Gaussian initial density profile} \label{sec:Gaussian}

The original (uniform-with-tail) initial particle density profile \eqp{ic} was chosen for analytical advantage: it is consistent with the Taylor approximation (Equations\ \ref{eq:forma}--\ref{eq:formd}) of the ODE model over a longer region near the origin and for a longer time.  Furthermore, the speed of a particle-gas wave is uniform, making the physics easier to analyze.  Here,  we explore some differences when the Gaussian initial profile,
\be
   \rhop(r, 0) = \rhopo \exp (-r^2/\sigmaG^2),
\ee
is used instead.  The definitions, \eqp{timedefs}--\eqp{Phidef}, for particle loading, dynamical time, and stopping time remain the same as before.  Recall that they are based on conditions at the origin.  The definition of the two-phase Jeans parameter, however, requires some comment.  Recall the definition \eqp{Jtuniform} used for the simulations:
\be
   \Jt \equiv \frac{\rhoto G R_0^2 (1 + \Phi_0)}{{\ci}^2}, \eql{JtOriginal_again}
\ee
where $R_0 = \Rcore + \sigmatail$ for the original profile.  We need to rewrite the clump radius $R_0$ so it has a common definition independent of profile shape.  The obvious way to do this (and in retrospect what should have been done in the first place) is to use the total particle mass.  For the original profile (where $\sigmatail  = 0.2 R_0$) the total particle mass is $M_\rmp = 4.017 \rhopo R_0^3$.  Hence we define a mass-effective clump radius
\be
   \Reff \equiv \left(\frac{M_\rmp}{4.017\rhopo}\right)^{1/3}. \eql{Reff_def}
\ee
and rewrite the definition of $\Jt$ as
\be
   \Jt \equiv \frac{\rhoto G \Reff (1 + \Phi_0)}{{\ci}^2}. \eql{newdef}
\ee
The Gaussian profile has $M_\rmp = \rhopo \pi^{3/2} \sigmaG^3$ which is substituted into \eqp{Reff_def} to obtain $\Reff$.  Code units for the Gaussian case are such that $\sigmaG = \ci = \rhogo = 1$.

Figure \ref{fig:t_collapse_Gaussian} shows collapse time as a function of $\Jt$.  Comparing this with Figure~\ref{fig:t_collapse_over_t_dyn}a for the uniform core case, one sees that while in the incompressible (i.e., sedimentation) limit the collapse time is the same (i.e., $\tcollapse/\tdyn$ very nearly equals $(4\pi\Stdyn)^{-1}$), as $\Jt$ increases, the Gaussian profile collapses more slowly.  This is probably because the mass function $M_\rmp(r)$ near the origin increases more slowly with $r$ for the Gaussian case.  A quadratic function of $\Jt$ was found necessary to fit the data:
\be
   \frac{\tcollapse}{\tdyn} = \max(F_\rmG(\Jt), 0.75), \eql{formula_gaussian}
\ee
with
\be
   F_\rmG(\Jt) \equiv \max[(4\pi\Stdyn)^{-1}, 0.75]\left[1 - 0.25 \Jt - 0.75 \Jt^2\right]. \eql{quadratic}
\ee
This fit is shown using dashed lines in Figure~\ref{fig:t_collapse_Gaussian}.  It should be noted that the Gaussian case also exhibited bouncing in certain cases.

\subsection{Numerical values for collapse time in a minimum-mass solar nebula}

Using formulas \eqp{formula} and \eqp{formula_gaussian}, collapse time was calculated for clumps located at the midplane of a minimum-mass solar nebula \citep{Hayashi1981} for which the parameters are:
\begin{align}
\ci  &= 10^5 (\varpi/\AU)^{-1/4}\ \cm\ \secinv,\\
\mathrm{gas\ scale\ height\ } H_\rmg &= \sqrt{2} \ci/\Omega_\mathrm{Kepler},\\
\mathrm{surface\ density\ } \Sigma_\rmg &= 1700 (\varpi/\AU)^{-3/2}\ \gm\ \cm^{-2},\\
\rhogo &= \Sigmag/(\pi^{1/2} H_\rmg).
\end{align}
The material density of solids is taken to be $\rho_\mathrm{material} = 3$ gm cm$^{-3}$.
Eight cases were selected: particle radius $a\in\left\{0.5\ \mm, 5\ \mm\right\} \times \varpi\in\left\{3\ \AU, 30\ \AU\right\}\times \Phi_0\in\left\{10, 100\right\}$; validity of the Epstein drag regime was verified for each case.  Results are shown in Figure~\ref{fig:t_collapse_numerical} as a function of normalized clump radius $\Reff/H_\rmg$.  Solid lines are for the uniform + tail initial profile and dashed lines are used for the Gaussian profile.  As $\Reff/H_\rmg \to 0$, the collapse time tends to the sedimentation time which is independent of the clump size.  Note that at a fixed radial location, the sedimentation time depends on the product $a \Phi_0$.  For $\Phi_0 = 100$, the clump radius has to be a few percent of the disk scale height for the Jeans parameter $\Jt$ to be sufficiently large that the collapse time is significantly reduced below the sedimentation time.  On the other hand, for $\Phi_0 = 10$ the clump radius has to be significant fraction of $H_\rmg$ for this to happen.  For the Gaussian initial profile (dashed lines)  the size required to overcome the sedimentation limit is even greater.  Currently, no mechanism is known that forms such large dense clumps.  We conclude that contraction of clumps containing mm to cm-sized particles will occur on the sedimentation timescale $\tsedsing = (4\pi G \rhopo \ts)^{-1}$.  Recall that this is a factor of $\pi$ smaller than earlier estimates \citep{Cuzzi_etal2008, Cuzzi_etal2010}.

\section{Layer-by-layer erosion due to turbulent strain}\label{sec:erosion}

Particle clumps will be subject to dispersive effects from within and without.  The former include turbulent gas motions within the clump, while the latter include the tidal force of the sun and ram pressure from a gas headwind \citep{Cuzzi_etal2008}.  
%
%
Here we are interested in the effect of eddies external to the clump which will induce a straining flow on the clump.  A process is described whereby, in time $\delta t$, the flow erodes a layer of thickness $\delta R$ from the periphery of the clump.

As a model problem, imagine a spherical clump of initial radius $R_0$ subject to a straining flow of characteristic strain rate $s$, for example, axisymmetric strain:
\be
   u_\rma = s x, \hskip0.4truecm v_\rma = -(s/2)y, \hskip0.4truecm w_\rma= -(s/2)z, \eql{axi_strain}
\ee
which is applied as the initial condition on the gas:
\be
   \bfu(\bfx, 0) = \bfu_\rma(\bfx),
\ee
as well as the boundary condition:
\be
   \bfu(\bfx, t) = \bfu_\rma(\bfx), \forall\,t > 0 \mathrm{\ as\ } |\bfx| \to \infty.
\ee
Define the characteristic velocity difference across the clump when its radius is $R$:
\be
   \Delta V(R) \equiv s R.
\ee
The important point to realize is that for particle loadings $\Phi \gg 1$, the flow \eqp{axi_strain} will be able to penetrate only a thin layer at the periphery of the clump.  To see this, note that the ``gas stopping time'' $\tg$ is
\be
   \tg = \ts/\Phi,
\ee
where $\ts$ is the particle stopping time and $\Phi$ is the particle loading.  In a time $\sim \tg$ from when the strain is turned on, the gas will come to rest throughout most of the clump and the particles will have moved very little.  Fresh gas entering the clump will come to rest in a layer of thickness
\be
   \delta R\sim \Delta V(R) \tg= s R \ts/\Phi. \eql{delta}
\ee
In homogeneous isotropic turbulence, the most strongly concentrated particles satisfy $s_\eta \ts \sim 1$, where $s_\eta$ is the strain rate at the Kolmogorov scale while the strain rate at scale $R$ (assumed to be in the inertial range is):
\be
   s(R) \sim \epsilon^{1/3} R^{-2/3},
\ee
where $\epsilon$ is the dissipation rate of turbulent kinetic energy per unit mass.
Substituting these facts into \eqp{delta} gives the thickness of the layer as:
\be
   \frac{\delta R}{R} \sim \left(\frac{\eta}{R}\right)^{2/3} \Phi^{-1},
\ee
which is indeed $\ll 1$ for $\eta/R \ll 1$ and $\Phi \gg 1$.
After a time $\ts$ from the initial condition, particles in the layer will start moving with the gas.
Imagine that, within this layer, a certain volume of fresh gas with speed $\Delta V(R)$ has combined with particles of mass loading $\Phi$ initially at rest.  Assume that particles and gas have small relative speed compared to their individual speeds.  Then conservation of total momentum gives the speed of the layer as:
\be
   U_\mathrm{layer} = \Delta V(R) / (1 + \Phi).
\ee
The layer will flow off the clump (i.e., be eroded) in time 
\be
   \delta t \sim R / U_\mathrm{layer}. \eql{layer_erosion_time} 
\ee
Taking the ratio of equations \eqp{delta} and \eqp{layer_erosion_time} and
replacing  the $\delta$ symbol with the infinitesimal ``$d$'' one obtains: 
\be
   \frac{dR}{dt} \sim -\frac{\Delta V(R)^2 \ts}{R\Phi (1 + \Phi)}. \eql{erosion_ode}
\ee

To estimate $\Delta V(R)$ for a turbulent flow consider the second-order velocity structure function defined as:
\begin{equation}
      D_{ij}(\bfr, \bfx, t) \equiv \left<\Delta u_i, \Delta u_j\right>,
\end{equation}
where angle brackets denote an ensemble average and $\Delta u_i$ is the velocity difference between two fixed points separated by $\bfr$: 
\begin{equation}
   \Delta u_i = u_i(\bfx + \bfr) - u_i(\bfx).
\end{equation}
The  Kolmogorov result for this is \citep[][p.193]{Pope2000}
\begin{equation}
   D_{ij} = C_2 (\epsilon r)^{2/3} \left(\frac{4}{3}\delta_{ij} - \frac{1}{3}\frac{r_i r_j}{r^2}\right),
\end{equation}
and experiments give $C_2 = 2.0$.
In the present context, it is better to think of $\bfx$ as Lagrangian point (i.e., a point following a gas particle) with the separation vector $\bfr$ slaved to the particle.  However, since a Lagrangian point uniformly samples the fluid volume, the distinction between a fixed $\bfx$ and Lagrangian $\bfx$ becomes immaterial once the ensemble average is taken.  Ideally, we would want $\bfx$ to follow a dust particle near the center of the clump; however, data for this is not available.
For the strain-rate, the longitudinal component, e.g. $D_{11}$, is relevant:
\begin{equation}
   \Delta V(R)^2 \sim D_{11} = C_2 (\epsilon R)^{2/3}.
\end{equation}

Finally, we account for the effect of the gravitational force of the clump in preventing erosion.  For this, we calculate the speed $V_\mathrm{levitate}$ of a gas flow which can just levitate a particle in the gravitational field of the clump.  The levitation speed is just the sedimentation speed, i.e., $V_\mathrm{levitate} = [G M(R)/R^2] \ts$.  If
\be
   \Delta V(R) < V_\mathrm{levitate}.  \eql{bound}
\ee
the strain flow will be unable to carry away particles.

From Equation~\eqp{erosion_ode}, the characteristic time for erosion is
\be
   t_\mathrm{erosion} \sim \frac{R_0^2 \Phi (1 + \Phi)}{\Delta V(R_0)^2 \ts}, \eql{erosion_time}
\ee
We assume, subject to \textit{a posteriori} checks, that the clump is in the sedimentation regime, and that the gravitational binding criterion \eqp{bound} is not satisfied.  Requiring that the sedimentation time be shorter than \eqp{erosion_time} gives:
\be
   \frac{4\pi\rhog  G R_0^2 \Phi^2 (1 + \Phi)}{\Delta V(R_0)^2} > 1. \eql{threshold}
\ee

A different way of thinking about clump distortion in a straining flow is to consider the ram pressure, following the treatment of Cuzzi et al. (\citeyear{Cuzzi_etal2008}, p.~1437) for the effect of a headwind.  The gradient of the ram pressure across the clump is $\rhog \Delta V^2 / R_0$.  This will distort the clump from being spherical, while the gravitational force (per unit volume) term, $\rhop G M_0 / R_0^2$,  will act to restore its shape ($M_0$ being the mass of the clump).  Requiring that the latter be larger than the former gives
\be
   \frac{4\pi \rhog G R_0^2\Phi^2}{3 \Delta V(R_0)^2} > 1, \eql{ram_pressure_criterion}
\ee
for the clump to resist distortion.  The ram pressure criterion \eqp{ram_pressure_criterion} is essentially the same as Equation (4) in \cite{Cuzzi_etal2008} or Equation (8) in \cite{Cuzzi_etal2010}.  Note that for $\Phi \gg 1$, the erosion criterion \eqp{threshold} is less restrictive than the ram pressure criterion \eqp{ram_pressure_criterion}.  Careful numerical experiments should be performed to test the above considerations.


\section{Concluding remarks}

This work presented solutions for the gravitational collapse of a spherical clump of particles in a gas as a function of the three governing parameters: the Stokes number $\Stdyn$, initial particle loading $\Phi_0$, and the two-phase Jeans number $\Jt$.
The last was shown to be the appropriate compressibility parameter in the context of particle collapse.  It is defined to be the classical Jeans number except that the speed of sound $\ci$ is replaced by an effective wave speed $\ci \phigo^{1/2} = \ci /(1 + \Phi_0)^{1/2}$ in a tightly coupled gas-particle medium.  Its use makes the results for collapse time generally insensitive to the initial particle loading $\Phi_0$ as a separate parameter although some details change.  On the other hand, use of the dynamical Mach number $\Madyn = t_\mathrm{acoustic}/\tdyn$ instead of $\Jt$ does not eliminate the $\Phi_0$ dependence.
It is important to emphasize the role of gas compressibility.  In particular, in the incompressible limit $\ci \to \infty$, the best one may expect (for particles of small Stokes numbers $\Stdyn$), no matter how high the particle density, is sedimentation rather than free-fall collapse.  Sedimentation leads to compaction on the slow timescale $t_\mathrm{sed,sing} = (4\pi \Stdyn)^{-1} \tdyn$.  This should be kept in mind when an incompressible flow solver is being used with particle gravity.  For larger particles, in particular when $\Stdyn \gtrsim 0.5$, the distinction between sedimentation and free fall blurs, and it is safe to assume incompressibility of the gas.
A simple ODE model was developed which captures essential features of the simulations and provides insight into clump oscillations (bouncing).

Formulae (fit to the simulations) were obtained for collapse time versus the governing parameters.  When applied to clumps consisting of mm or cm-sized particles in a minimum-mass solar nebula, these formulae suggest that their gravitational contraction occurs on the self-sedimentation timescale.  The collapse time is reduced significantly below the sedimentation time only for clumps whose size is a significant fraction of the disk scale height and currently no mechanism is known that can produce such clumps.  The sedimentation time is independent of clump size and therefore cannot  directly set the initial mass function.

A dense particle clump in a protoplanetary disk will be subject to various dispersive effects and we considered one of these, namely, stripping of the clump by turbulent strain.  For planetesimal formation it is required that collapse time be shorter than dispersal time.  This provides an additional constraint on the critical clump size and particle concentration.  Finally, to develop the initial mass function for planetesimals, the rate of occurrence of clumps satisfying these constraints is required.  For the case of preferential concentration by turbulence, work along these lines can be found in \cite{Cuzzi_etal2010} and \cite{Hopkins2014}.  The former uses a cascade model; however, \cite{Pan_etal2011} based on their higher-Re simulations questioned whether the self-similarity it assumes actually holds and the cascade multipliers it uses are valid at large scales in the inertial range.  This concern had already been identified as worth further study by \cite{Cuzzi_etal2010}, and ongoing work has indeed shown that the concentration cascade is not scale invariant, at least on the large scales of most interest to the nebula problem. In particular, \cite{Cuzzi_etal2014} have used even higher Re simulations \citep{Bec_etal2010} to measure the scale dependence of the multiplier functions that determine the outcomes of the cascades, and begun to construct new cascades based on the new results. While those particles having stopping time equal to the Kolmogorov time (which were the only focus of interest previously) are less rapidly and strongly concentrated than before, particles just a few times larger are still concentrated significantly. There will be implications for the probability distributions of dense clumps, for planetesimal formation statistics, for chondrule size distributions, and for the most promising combinations of particle size, nebula gas density, and turbulent intensity.  These will become clear only after a new set of complete scale-dependent cascade models has been run using the new multiplier distributions.
 

%
The present calculations were not continued past the point of singularity formation at the origin.  In the star-formation literature this is referred to as point-mass formation (PMF).  In the present context, the singularity can be ameliorated by removing two assumptions.  Removing the constraint of spherical symmetry would shift the singularity to a caustic surface near the origin.  Second, even at the caustic surface, particle streams traveling in opposite directions would (in the absence of collisions) interpenetrate without causing a singularity in particle density.  The two-fluid model, however, averages the velocity of the two streams to zero and a singularity in particle density occurs.  It may be of interest to continue the calculation (even in the spherically symmetric case) past the singularity using velocity moment methods \citep[e.g.,][]{Chalons_etal2012}.

Isothermal conditions were assumed throughout.  If heat generated due to drag remains trapped (due to radiation absorption by small particles) then the retarding effect of gas pressure will be enhanced.  Although we have not done so, the adiabatic limit is straightforward to treat.

%

\begin{center}\textbf{Acknowledgements}\end{center}

We are grateful to the internal reviewers, Drs. Anthony Dobrovolskis, Terry Holst, and Alan Wray, for their helpful comments.
We are grateful to Profs. Marc Massot (Ecole Centrale) and Ali Mani (Stanford Univ.) for enlightening us on the defect of the two-fluid model at crossing trajectories.  We thank the referee, Dr. Stuart Weidenschilling (Planetary Science Institute), for suggestions to improve the paper.

\bibliographystyle{apj}
\bibliography{paper}

\clearpage

\clearpage
\clearpage
\clearpage
\begin{figure}
\begin{center}
\includegraphics[width=4.5in]{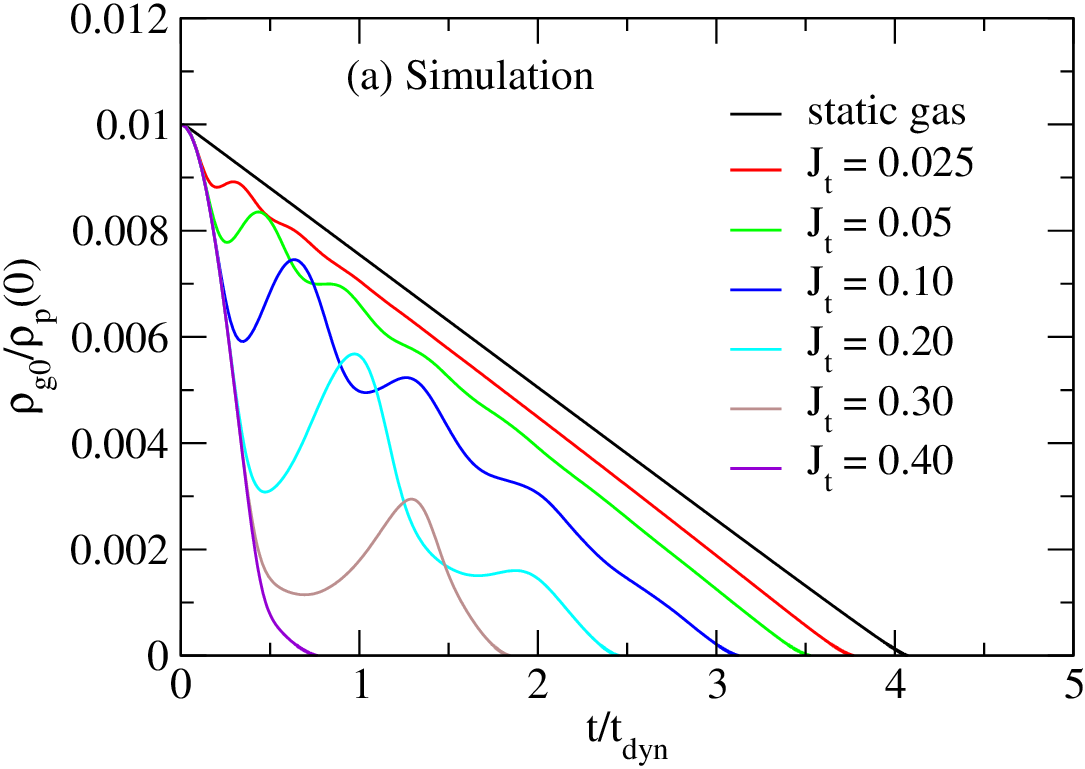}
\vskip 1.1truecm
\includegraphics[width=4.5in]{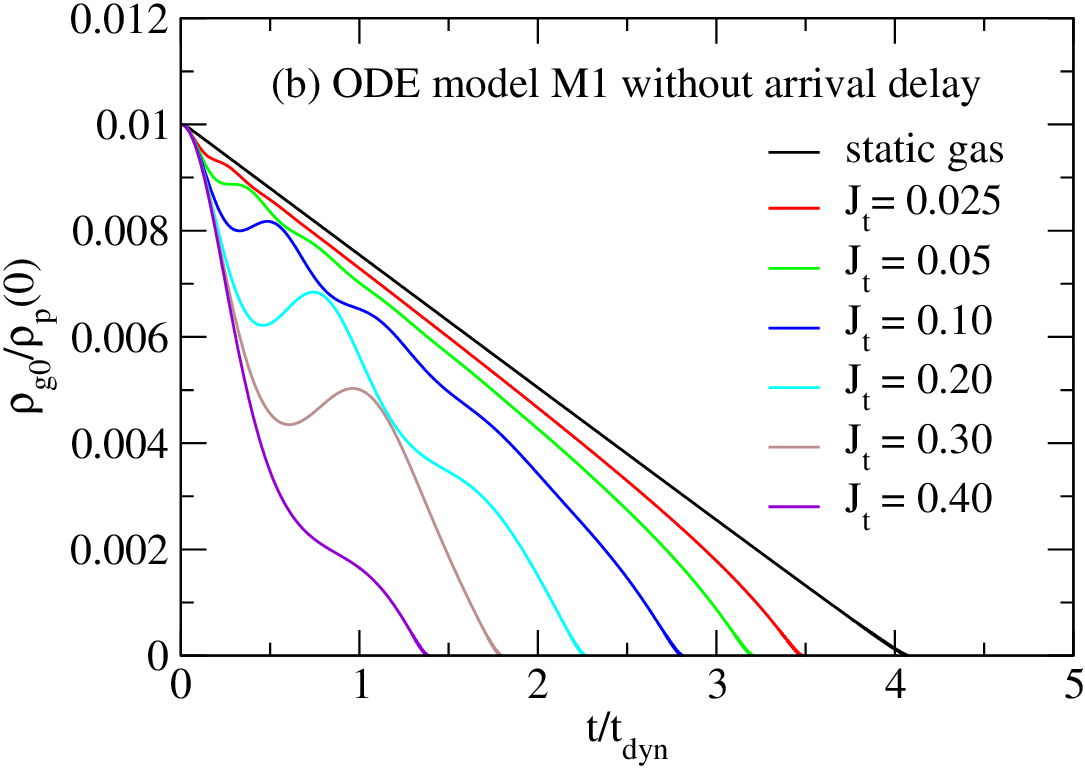}
\end{center}
Figures \ref{fig:specific_volume}a and b.  For Figure \ref{fig:specific_volume}c and complete caption please see the next page.
\end{figure}

\begin{figure}
\begin{center}
\includegraphics[width=4.5in]{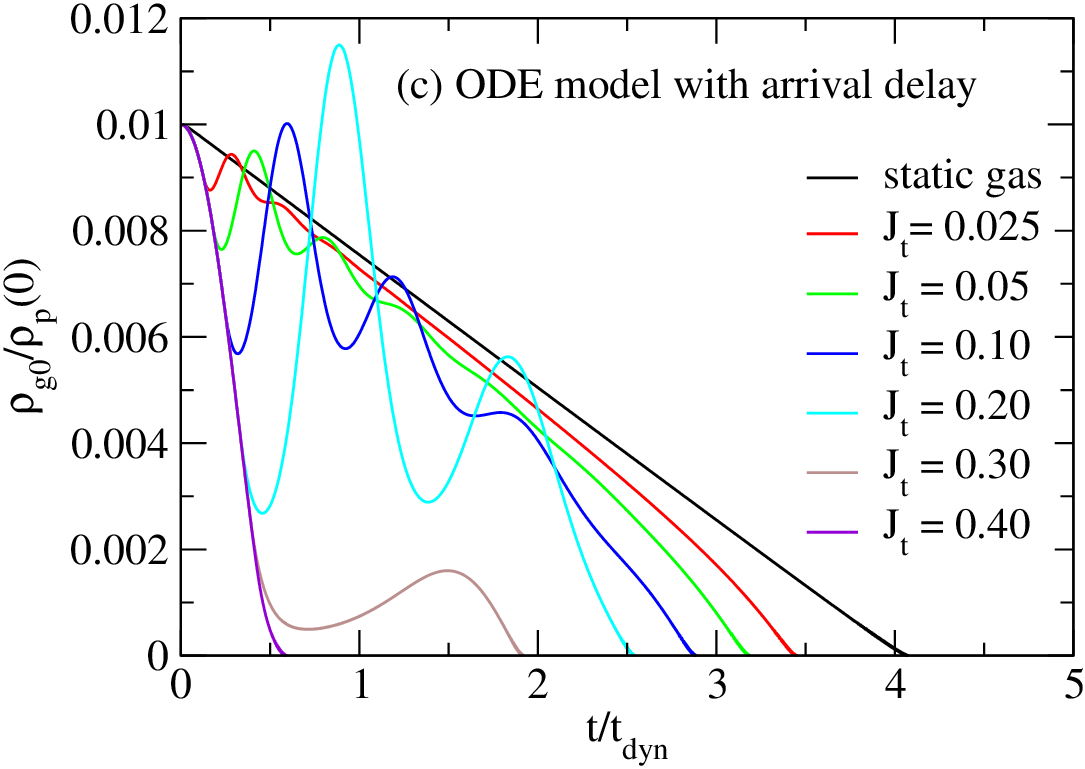}
\end{center}
\caption{Specific particle volume at the origin (normalized by the specific volume, $1/\rhogo$, of the ambient gas).  (a) Simulations. (b) ODE model M1 without arrival delay ($\frg = 0.43$).  (c) ODE model with arrival delay ($\frg = 0.4579$). $\Phi_0 = 100, \Stdyn = 0.02$.}
\label{fig:specific_volume}
\end{figure}
\begin{figure}
\begin{center}
\includegraphics[width=5.5in]{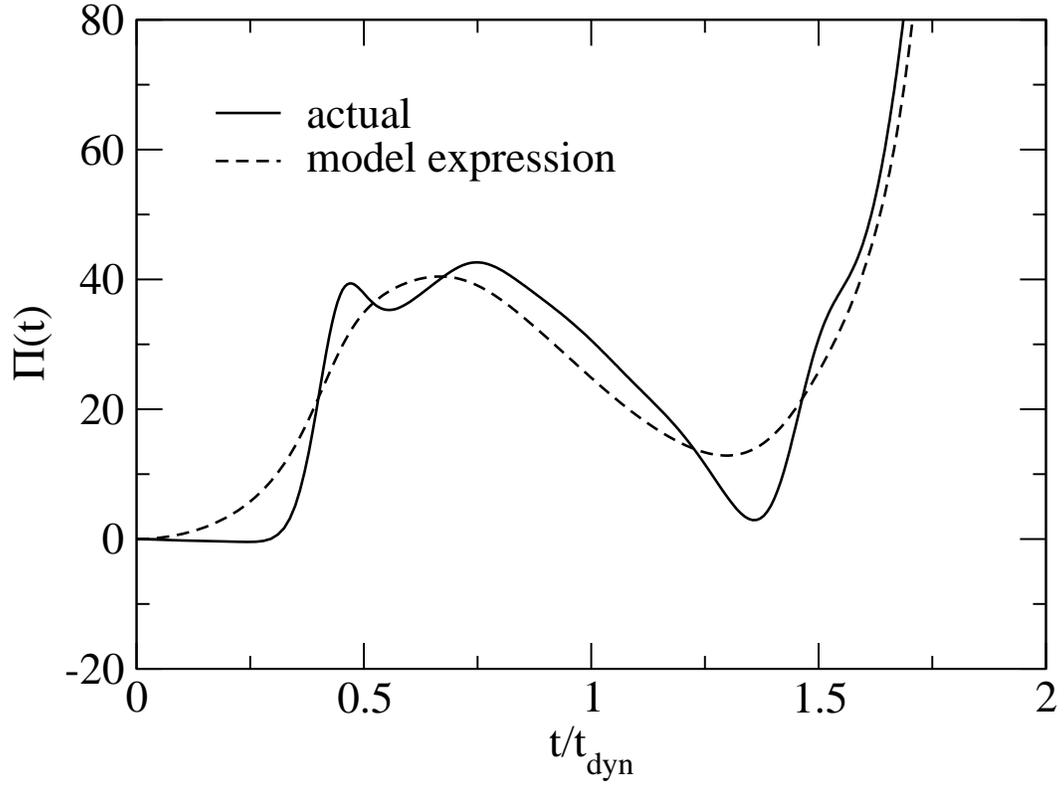}
\end{center}
\caption{Pressure gradient term $\Pi(t)$ as evaluated in the simulation.  Solid: Actual value from its definition \eqp{pg_term}; Dashed: M1 model expressions \eqp{pg_model} and \eqp{rg_model} using $\frg = 0.5$
($\Phi_0 = 100, \Jt = 0.30, \Stdyn = 0.02$).}
\label{fig:pg_term}
\end{figure}
\begin{figure}
\begin{center}
\includegraphics[width=4.5in]{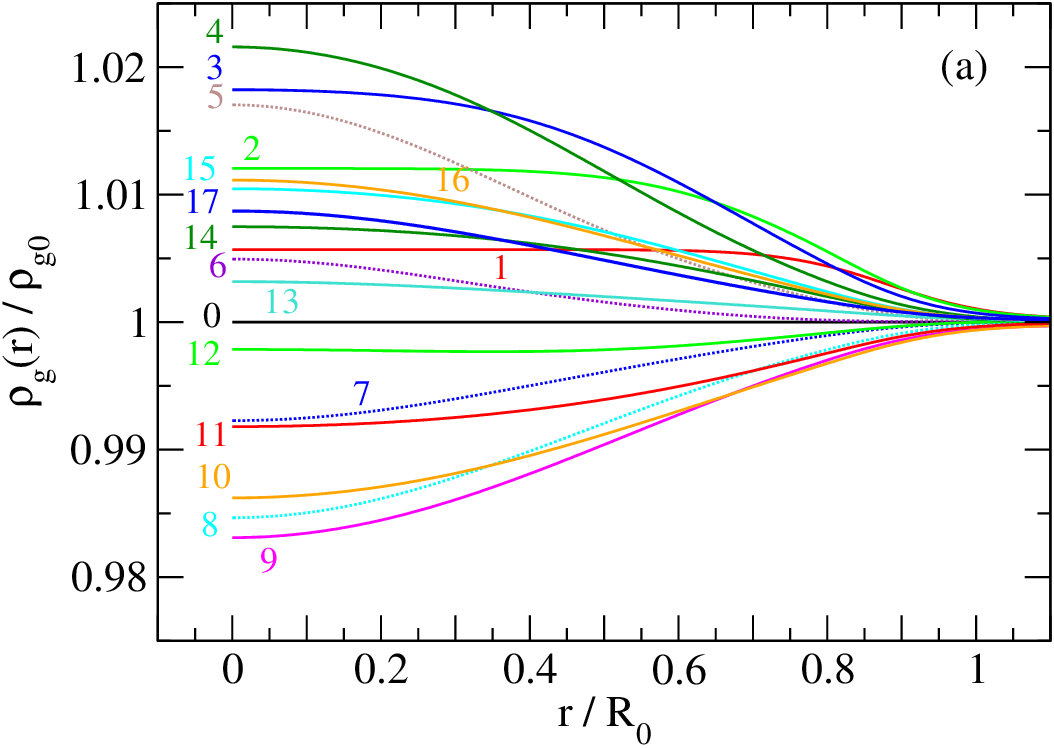}
\vskip1.5truecm
\includegraphics[width=4.5in]{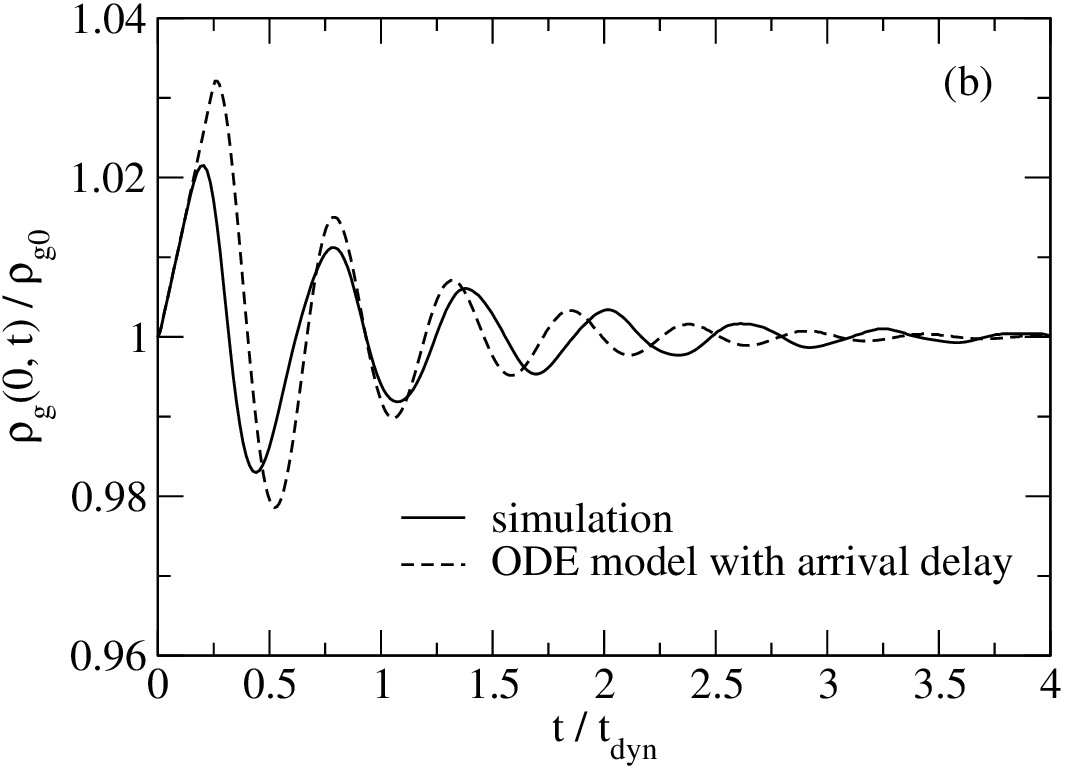}
\end{center}
\caption{Damped standing wave for the the case where gravity is turned off for $t / \tdyn > 0.01$.  (a) Gas density profiles at various instants.  Note that in frames 0 to 3 the inner part of the profiles remains flat as the pressure gradient effect propagates inward.  The various instants are separated by $\Delta t / \tdyn = 0.05$.  (b) Gas density at the origin.
Parameters: $\Phi_0 = 100, \Jt = 0.10, \Stdyn = 0.02$.}
\label{fig:standing_wave}
\end{figure}
\begin{figure}
\begin{center}
\includegraphics[width=4.5in]{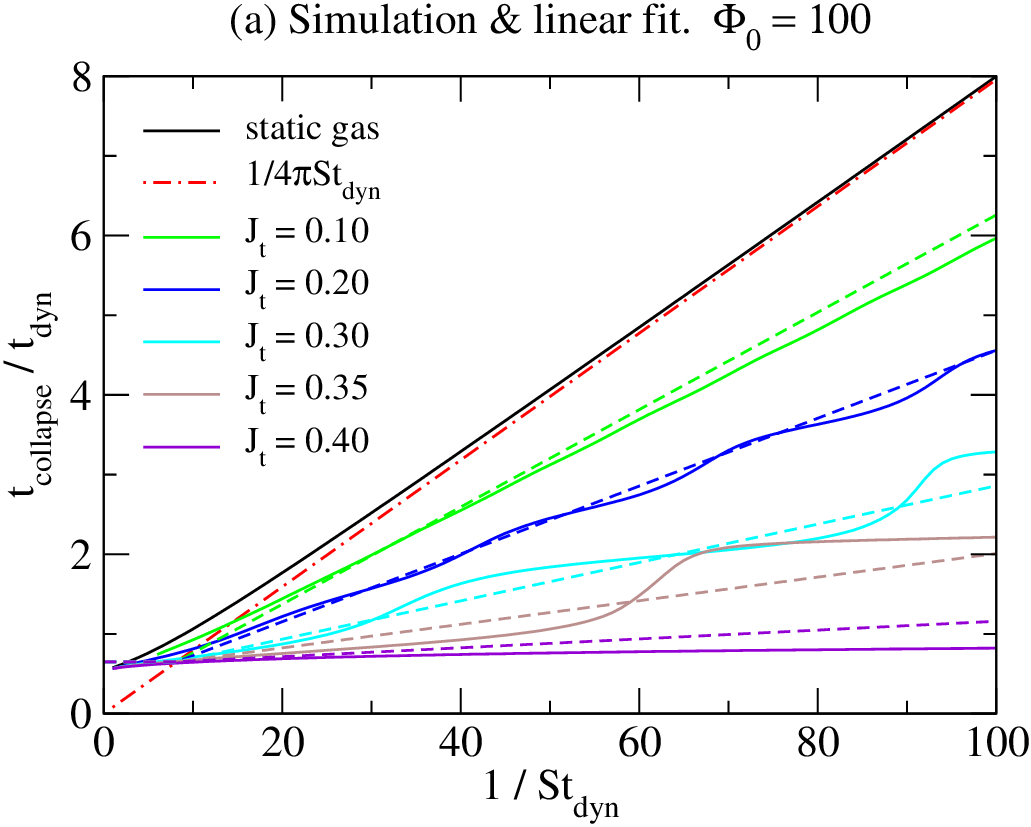}
\vskip 1.1truecm
\includegraphics[width=4.5in]{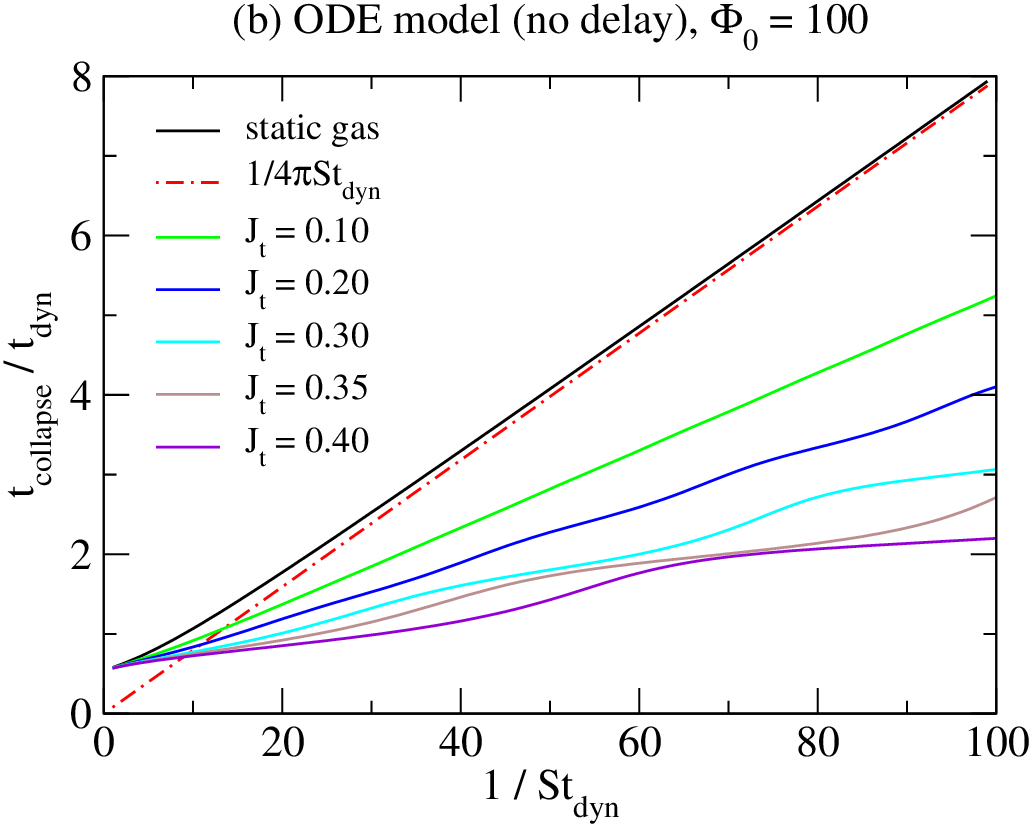}
\end{center}
Figures \ref{fig:t_collapse100}a and b.  For Figure \ref{fig:t_collapse100}c and complete caption please see the next page.
\end{figure}
\begin{figure}
\begin{center}
\includegraphics[width=4.5in]{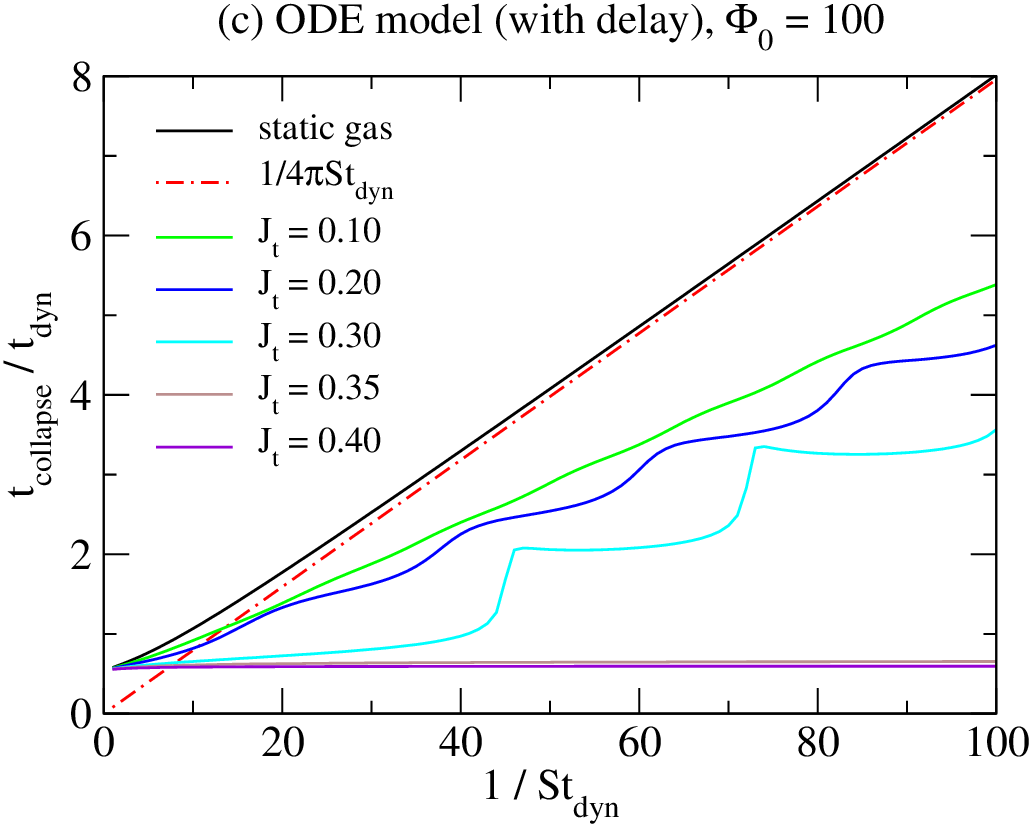}
\caption{Collapse time versus $\Stdyn^{-1}$ for various $\Jt$.  $\Phi_0 = 100$.  Note: Each curve has about 100 points and each point represents a separate simulation.  The dashed lines in Figure \ref{fig:t_collapse100}a represent the linear fit, Equation \eqp{formula}.}
\label{fig:t_collapse100}
\end{center}
\end{figure}
\clearpage
\begin{figure} \begin{center}
\includegraphics[width=4.5in]{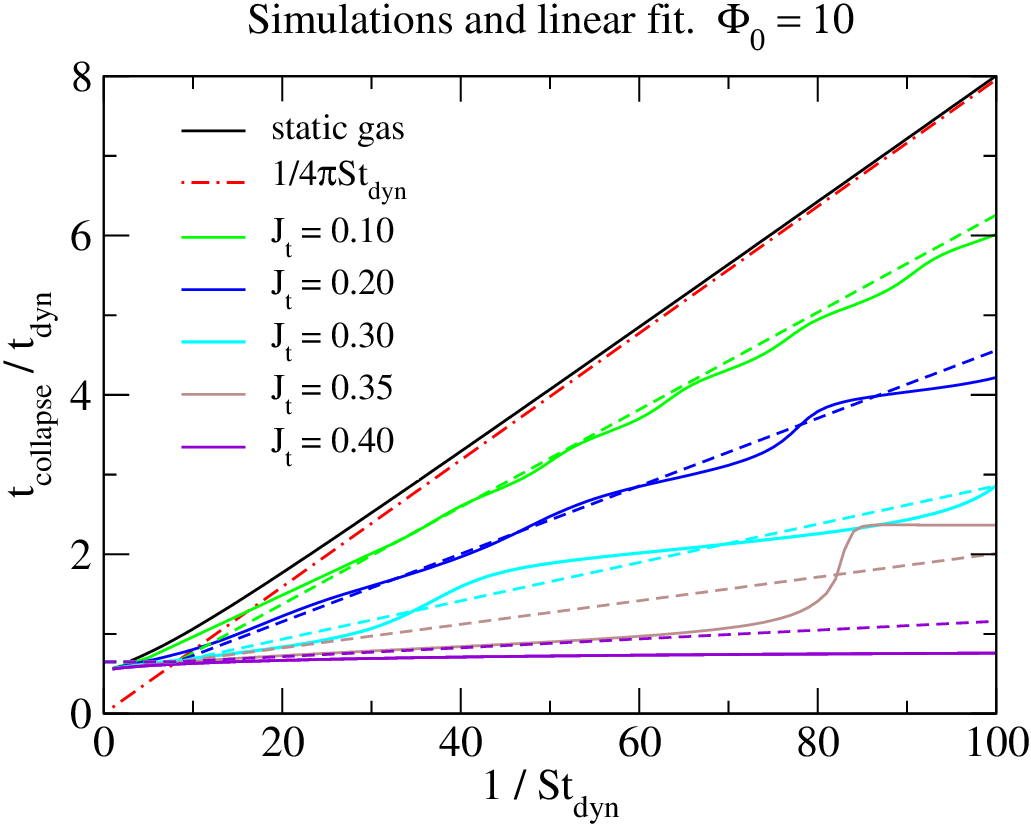}
\end{center}
\caption{Collapse time versus $\Stdyn^{-1}$ for various $\Jt$.  $\Phi_0 = 10$.  Solid: simulations; dashed: linear fit \eqp{formula}.  Every solid line has about 100 points and each point corresponds to a separate simulation.}
\label{fig:t_collapse10}
\end{figure}
\begin{figure}\begin{center}
\includegraphics[width=4.5in]{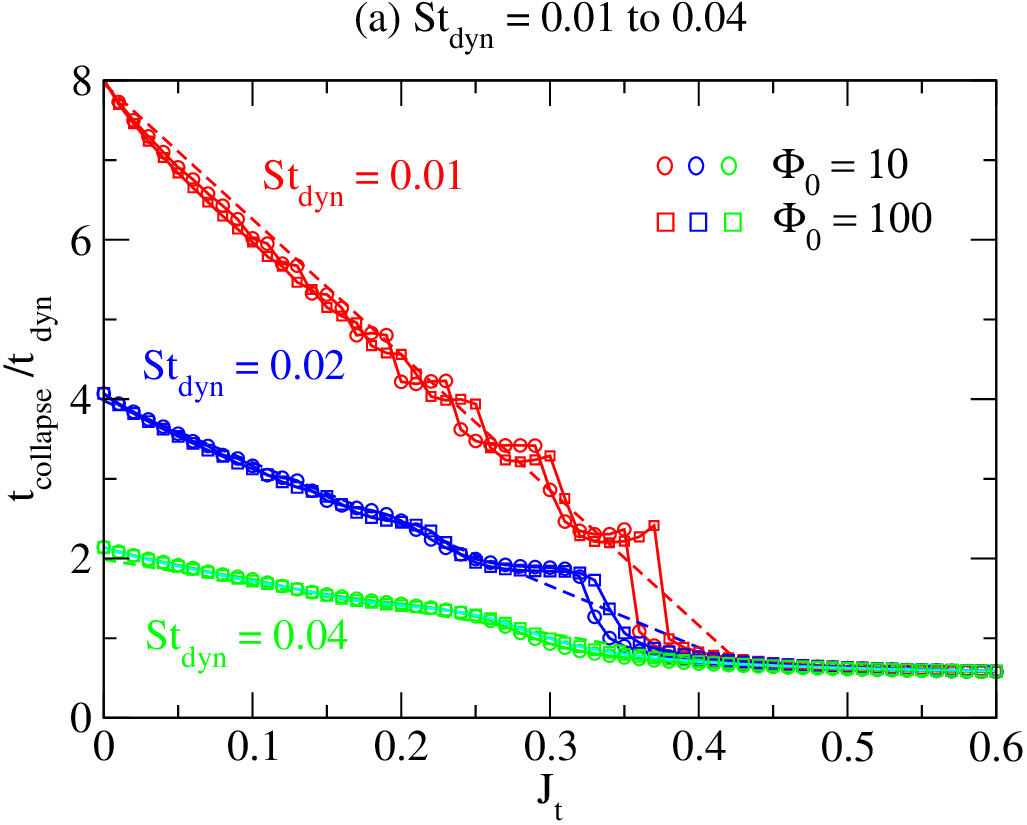}
\vskip 1.0truecm
\includegraphics[width=4.5in]{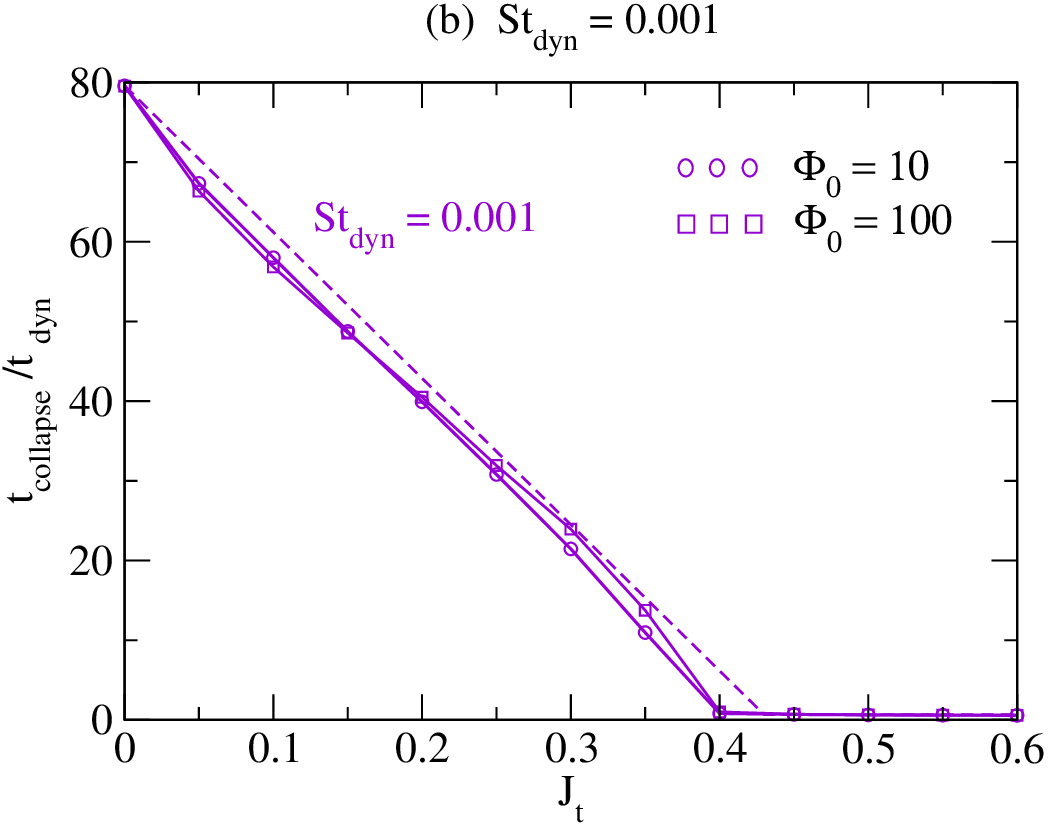}
\end{center}
\caption{Collapse time versus $\Jt$ for various Stokes numbers.  Each symbol represents a different simulation.  Dashed lines show the result of the linear fit, Equation~\eqp{formula}.}
\label{fig:t_collapse_over_t_dyn}
\end{figure}
\begin{figure}\begin{center}
\includegraphics[width=4.5in]{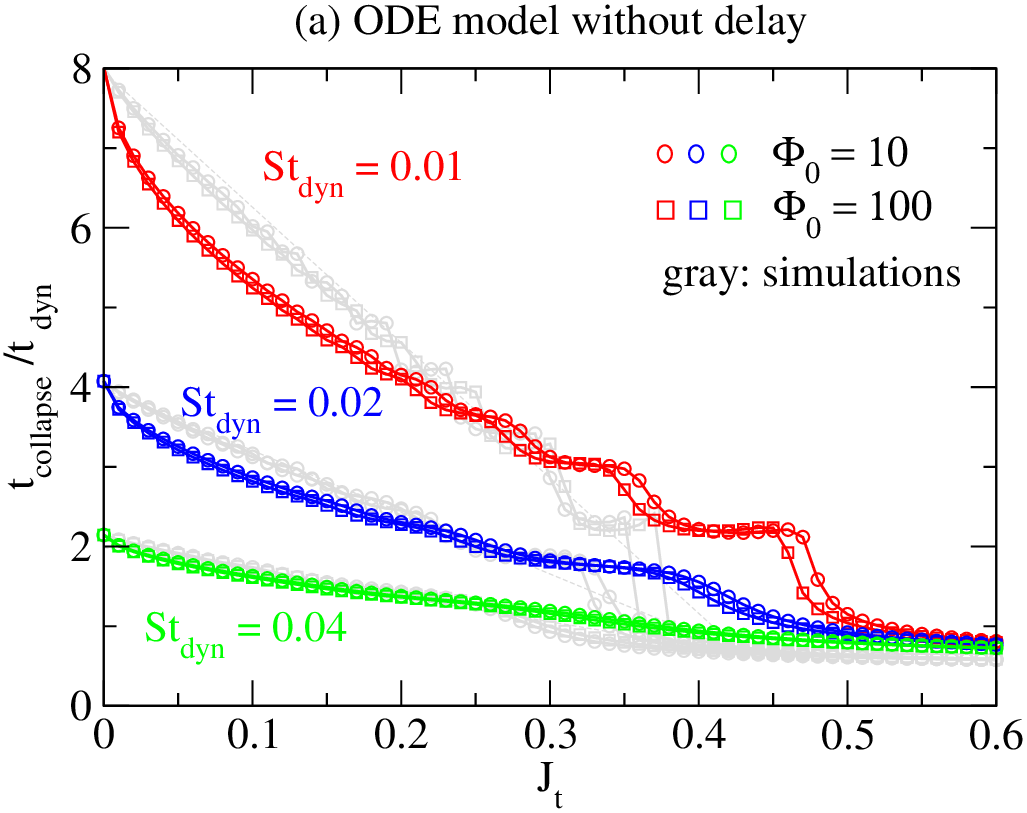}
\vskip 1.2truecm
\includegraphics[width=4.5in]{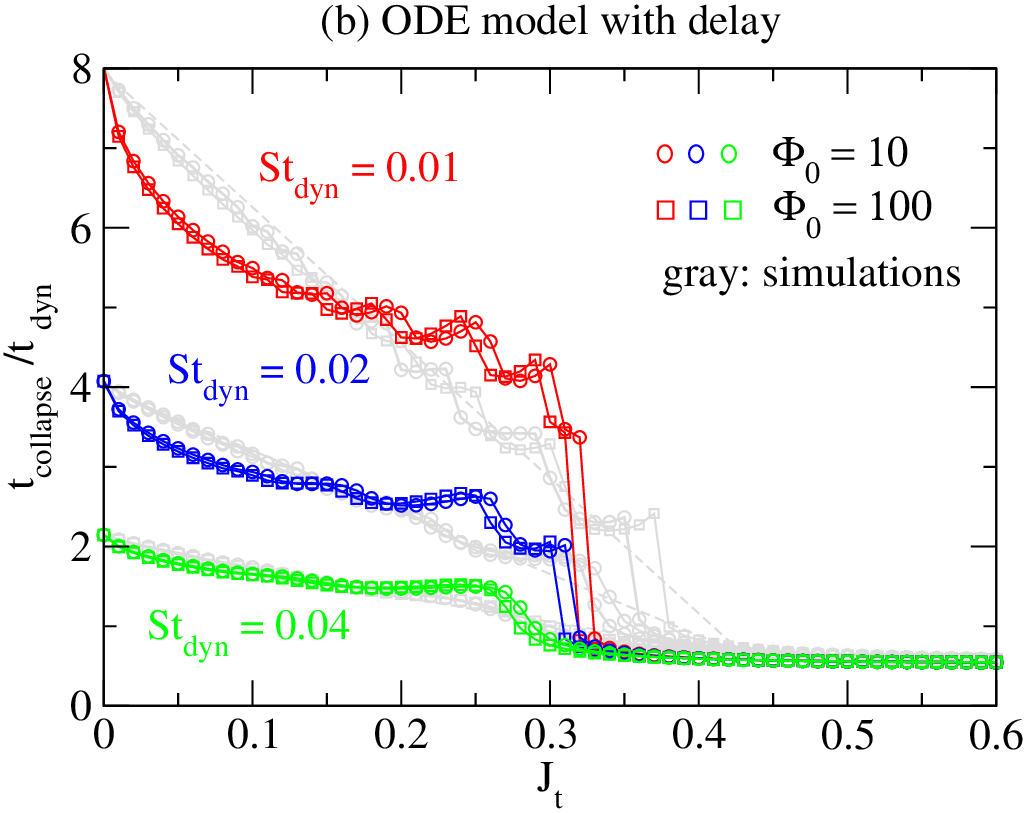}
\end{center}
\caption{Collapse time versus $\Jt$ given by the ODE models.  Each symbol represents a separate run.}
\label{fig:t_collapse_versus_Jt_ODE}
\end{figure}
\begin{figure}\begin{center}
\includegraphics[width=4.5in]{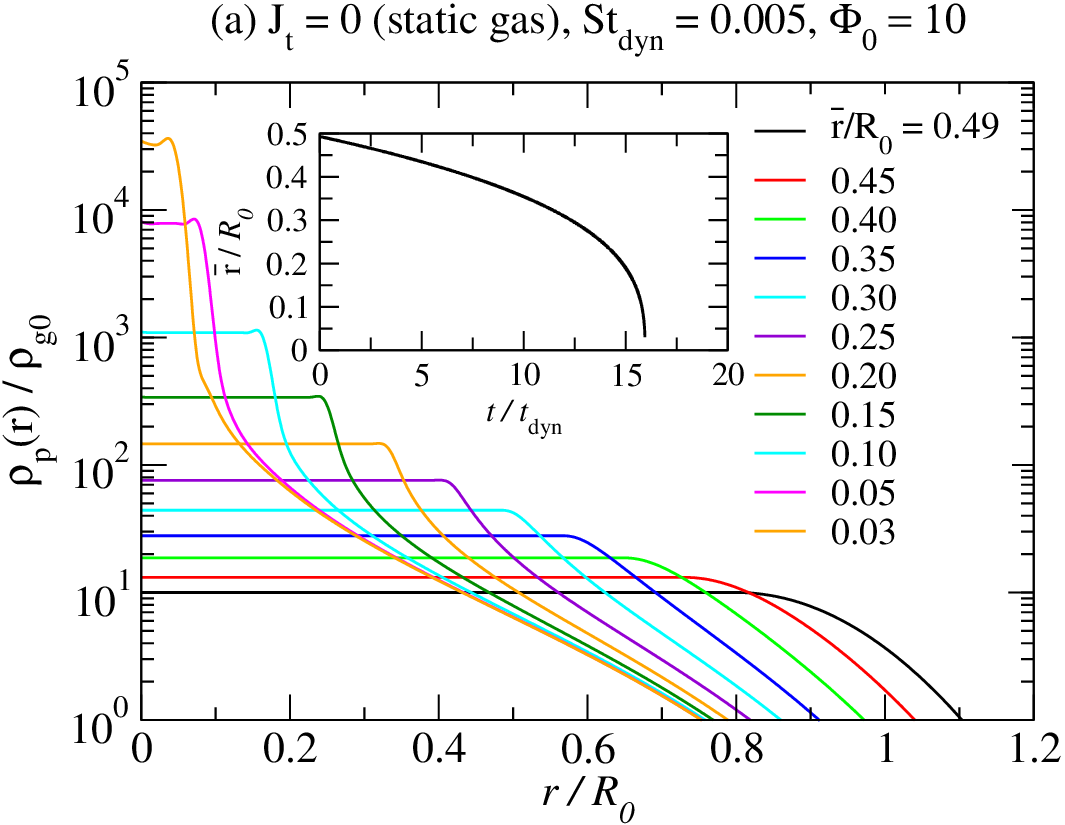}
\vskip 1.5truecm
\includegraphics[width=4.5in]{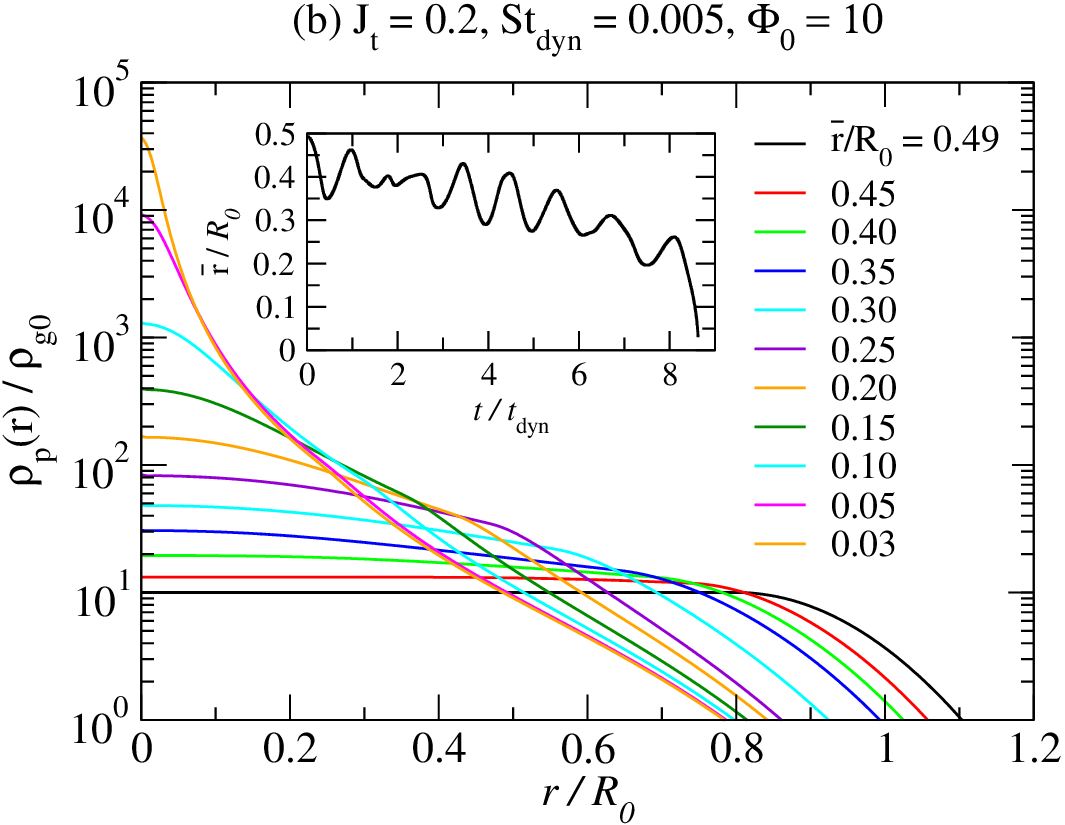}
\end{center}
Figures \ref{fig:profiles}a and b.  For Figures \ref{fig:profiles}c and d please see the next page.
\end{figure}
\clearpage
\begin{figure}\begin{center}
\includegraphics[width=4.5in]{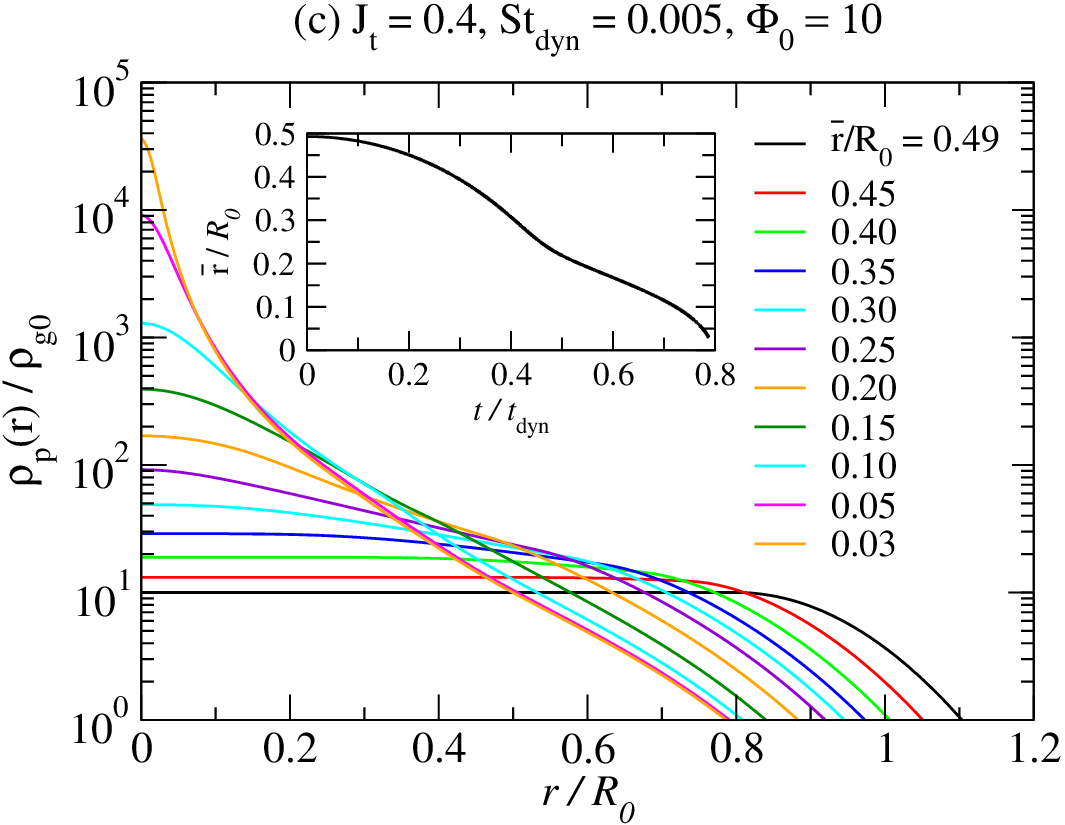}
\vskip 1.5truecm
\includegraphics[width=4.5in]{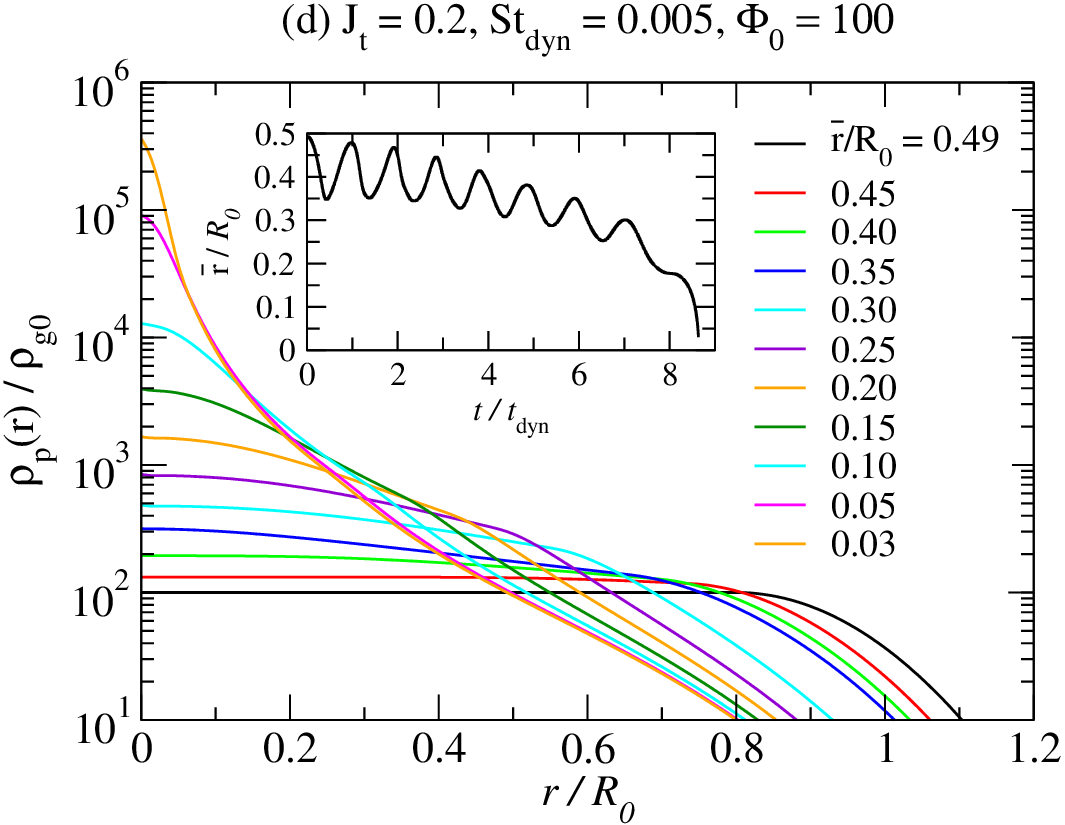}
\end{center}
Figures \ref{fig:profiles}c and d.  For Figures \ref{fig:profiles}e and f and the caption please see the next page.
\end{figure}
\clearpage
\begin{figure}\begin{center}
\includegraphics[width=4.5in]{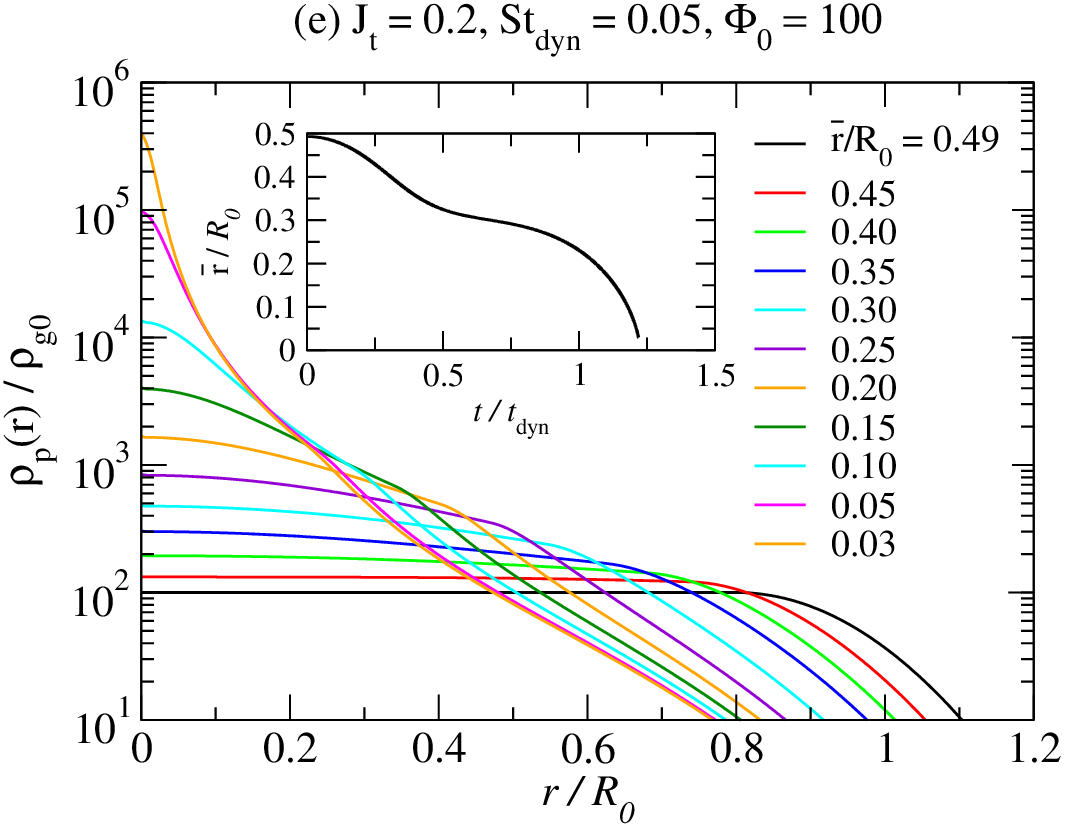}
\vskip 1.5truecm
\includegraphics[width=4.5in]{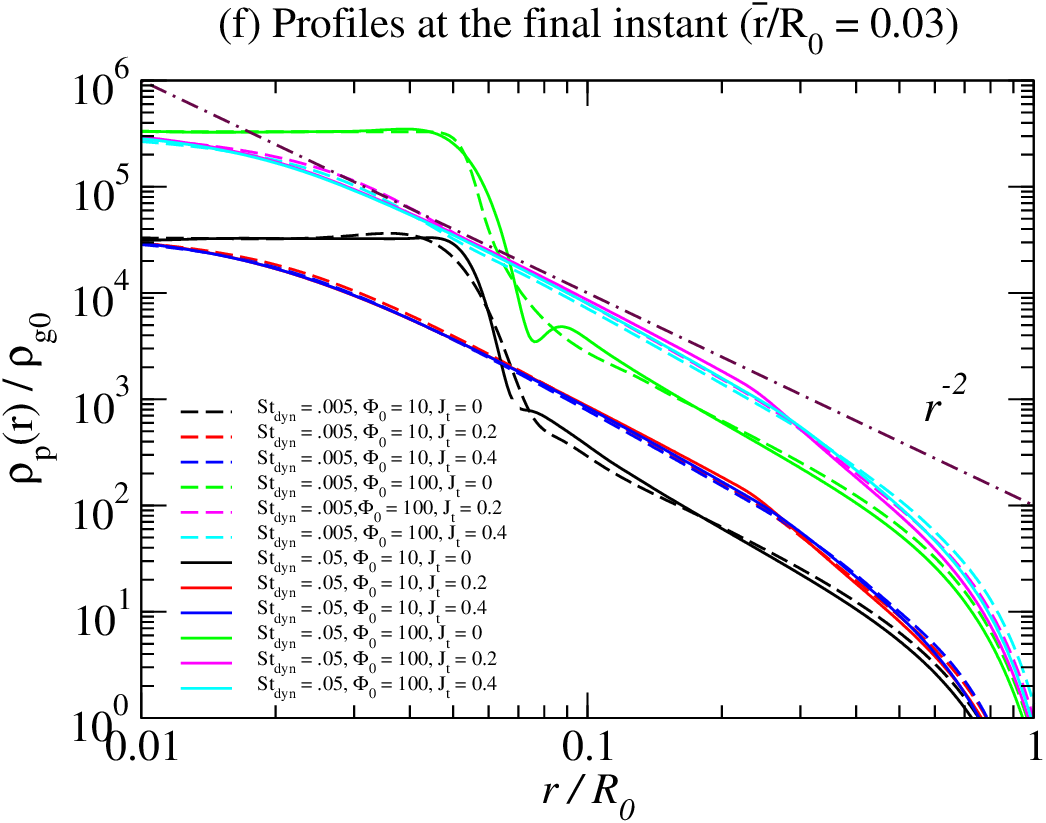}
\end{center}
\caption{Time evolution of particle density profiles for various cases.}
\label{fig:profiles}
\end{figure}
\clearpage
\begin{figure}\begin{center}
\includegraphics[width=5.5in]{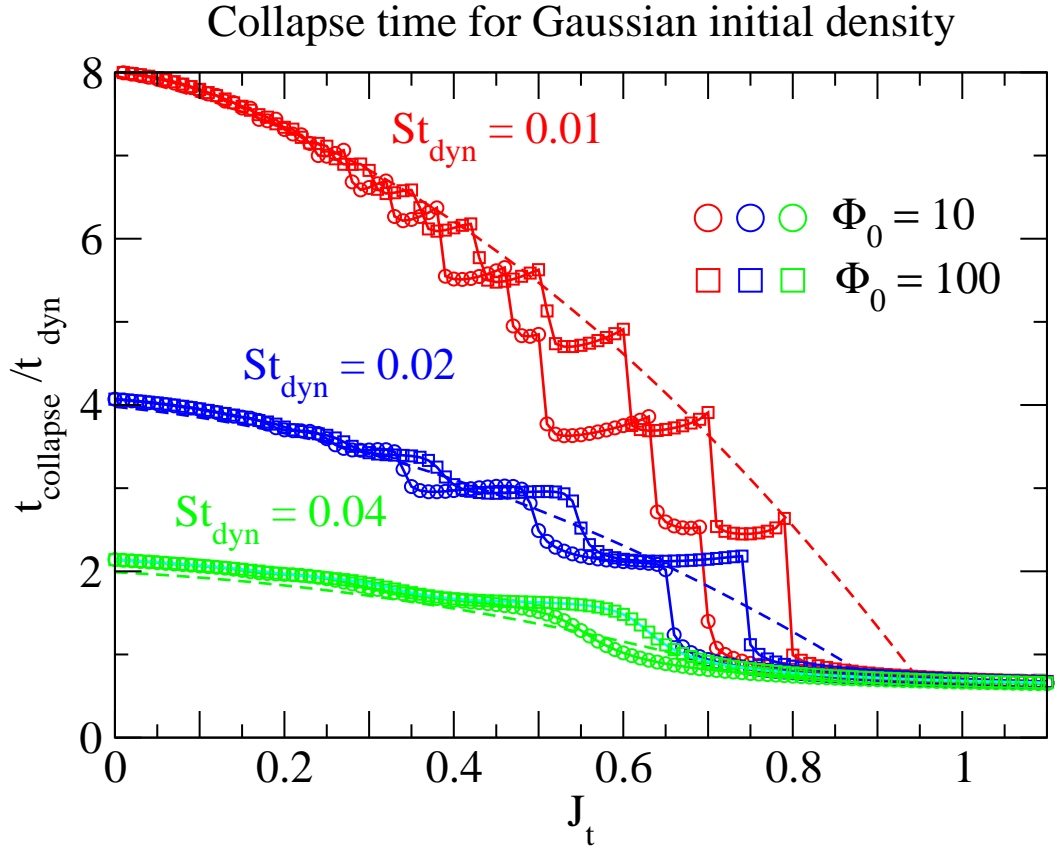}
\caption{Collapse time for an initially Gaussian profile of particle density.  Each symbol represents a separate simulation.  The dashed lines represent the quadratic fit, Equation \eqp{quadratic}.}
\label{fig:t_collapse_Gaussian}
\end{center}
\end{figure}
\clearpage
\begin{figure}
\begin{center}
\includegraphics[width=5.5in]{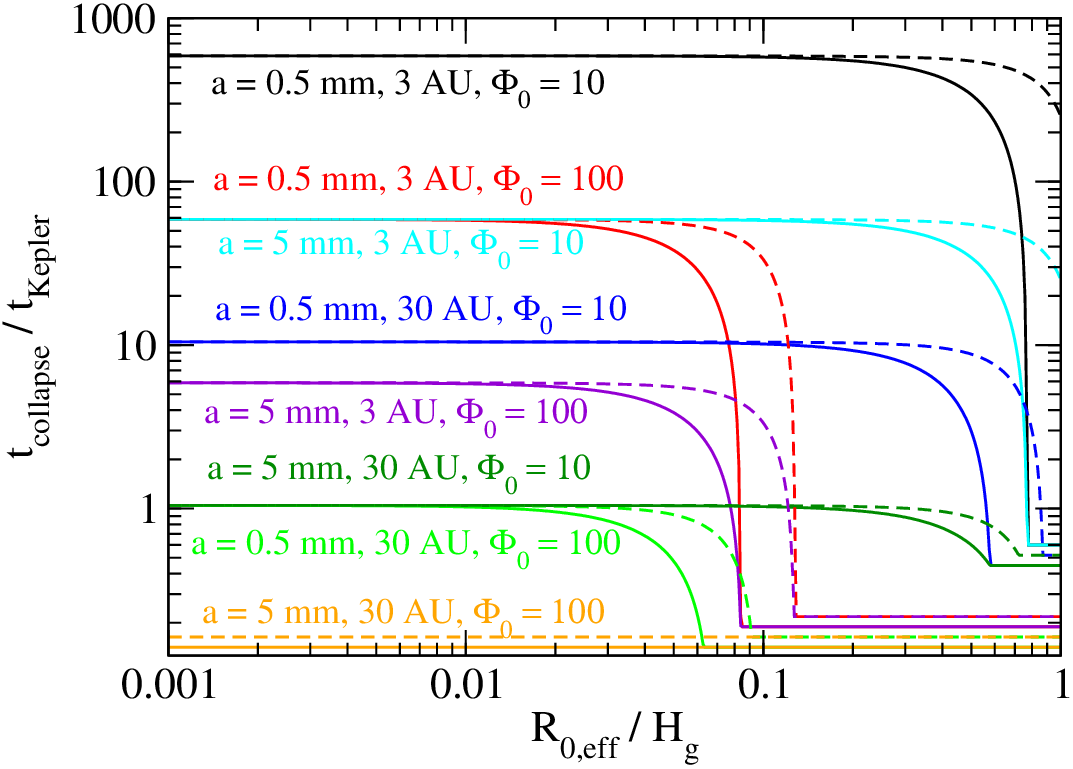}
\caption{Collapse time versus normalized clump radius $\Reff/H_\rmg$ in a minimum mass solar nebula using the fits \eqp{formula} and \eqp{formula_gaussian}.  $a$ is the particle radius and $H_\rmg$ is the gas scale height of the disk.  Solid lines: Uniform $+$ tail initial density profile; Dashed lines: Gaussian initial density profile.}
\label{fig:t_collapse_numerical}
\end{center}
\end{figure}
\end{document}